\documentclass[fleqn,usenatbib]{mnras}
\usepackage{newtxtext,newtxmath}
\usepackage[T1]{fontenc}
\usepackage{float}

\DeclareRobustCommand{\VAN}[3]{#2}
\let\VANthebibliography\thebibliography
\def\thebibliography{\DeclareRobustCommand{\VAN}[3]{##3}\VANthebibliography}

\usepackage{graphicx}	% Including figure files
\usepackage{amsmath}	% Advanced maths commands
\usepackage{threeparttable}
\usepackage{xcolor}

\graphicspath{{figures/}}

\title[TXS 2005+403 multiple imaging formed by scattering]{Multiple imaging of the quasar 2005+403 formed by anisotropic scattering}
\author[Koryukova et al.]{\parbox{\textwidth}{
T.~A. Koryukova,$^{1}$\thanks{E-mail: tatyana.koryukova@gmail.com}
A.~B. Pushkarev,$^{2,1}$
S.~Kiehlmann,$^{3,4}$
A.~C.~S.~Readhead$^{5}$
}
\vspace{0.4cm}\\
\parbox{\textwidth}{
$^1$Lebedev Physical Institute of the Russian Academy of Sciences, Leninsky prospekt 53, 119991 Moscow, Russia\\
$^2$Crimean Astrophysical Observatory, Nauchny 298409, Crimea, Russia\\
$^3$Institute of Astrophysics, Foundation for Research and Technology-Hellas, GR-70013 Heraklion, Greece\\
$^4$Department of Physics and Institute of Theoretical and Computational Physics, University of Crete, 70013 Heraklion, Greece\\
$^5$Owens Valley Radio Observatory, California Institute of Technology, Pasadena, CA 91125, USA
}}

\date{Accepted 2023 October 2. Received 2023 September 30; in original form 2023 August 24}

\pubyear{2023}

\begin{document}
\label{firstpage}
\pagerange{\pageref{firstpage}--\pageref{lastpage}}
\maketitle

% Abstract of the paper
\begin{abstract}
We report on the low Galactic latitude ($b=4\fdg3$) quasar 2005$+$403, the second active galactic nuclei, in which we detected a rare phenomenon of multiple imaging induced by refractive-dominated scattering. The manifestation of this propagation effect is revealed at different frequencies ($\lesssim8$~GHz) and epochs of VLBA observations. The pattern formed by anisotropic scattering is stretched out along the line of constant Galactic latitude with a local $\mathrm{PA}\approx40^\circ$ showing one-two sub-images, often on either side of the core. Analysing the multi-frequency VLBA data ranging from 1.4 to 43.2~GHz, we found that both the angular size of the apparent core component and the separation between the primary and secondary core images follow a wavelength squared dependence, providing convincing evidence for a plasma scattering origin for the multiple imaging. Based on the OVRO long-term monitoring data at 15~GHz obtained for 2005$+$403, we identified the characteristic flux density excursions occurred in April-May 2019 and attributed to an extreme scattering event (ESE) associated with the passage of a plasma lens across the line of sight. Modeling the ESE, we determined that the angular size of the screen is 0.4~mas and it drifts with the proper motion of 4.4~mas~yr$^{-1}$. Assuming that the scattering screen is located in the highly turbulent Cygnus region, the transverse linear size and speed of the lens with respect to the observer are 0.7~AU and 37~km~s$^{-1}$, respectively.
\end{abstract}


%\newpage
%\input{text_of_article}

\section{Introduction}
\label{s:intro}

If a bright and compact background radio source is observed at low Galactic latitudes and low frequencies (below 8~GHz), its radio emission can be scattered by especially dense \citep[with free-electron density $n_e \approx10^{4-5}$~cm$^{-3}$, e.g.][]{Clegg_1998, Pushkarev2013} and inhomogeneous regions of the interstellar medium (ISM) of the Milky Way. The large-scale plasma associations \citep[$10^{13}-10^{14}$~cm, ][]{Romani1987} in the Galactic thin disk are also called plasma lenses because of their ability to refract radio waves. This phenomenon gives rise to remarkable observed effects, including (i) formation of caustic surfaces on a light curve, where the apparent brightness of the background source may  experience a significant increase, (ii) creation of multiple images and angular position wandering of the background radio source \citep{Clegg_1998}. Previous studies have explored the potential origins of these strong lensing effects within the ISM. For instance, \cite{Cordes1988} have suggested that strong lensing in the ISM occurs for the lines of sight through the edge of a Galactic outflow, such as a supernova remnant. \cite{Clegg1988} have demonstrated that ionized shock fronts viewed edge-on can produce notable refraction effects at radio frequencies. However, the exact nature of scattering lenses within the Galaxy remains an open question, warranting further investigation.

Active galactic nuclei (AGN) observed at gigahertz frequencies using very long baseline interferometry (VLBI) typically manifest compact radio structure with one-sided core-jet morphology 
due to relativistic beaming 
\citep[e.g.][]{Kellermann07}. However, \cite{Pushkarev2013}, in their analysis of multi-epoch VLBA observational data of the quasar 2023$+$335 at 15.4~GHz, made a remarkable discovery. They observed highly unusual changes in the brightness distributions of the quasar during certain epochs, wherein additional bright emitting regions appeared. These peculiar structural variations were induced by scattering and represented the first successful observational detection of the theoretically predicted multiple imaging effect of a radio source as a result of refractive-dominated scattering in the ISM. This phenomenon also coincided temporally with the presence of a caustic surface on the 15~GHz Owens Valley Radio Observatory (OVRO) light curve, indicating that the source was undergoing an extreme scattering event (ESE) caused by the refraction of radio waves in the ISM \citep{Fiedler1987, Romani1987, Clegg_1998}. 

\cite{Vedantham2017} analysed the multiple U-shaped patterns detected in the light curve of the source PKS~1413$+$135 at different frequencies. They critically examined the hypothesis of an intermediate plasma lens as the cause of an ESE, whose $\lambda^2$-dependence (where $\lambda$ is the observing wavelength) is inconsistent with the observed achromatic variability pattern. \cite{Vedantham2017b} discussed that these patterns arises from gravitational milli-lensing and, more recently, \cite{Readhead21} concluded that the gravitational lens is hosted in an intervening spiral galaxy. Thus, an observed symmetric variability pattern in light curves can not be considered sufficient evidence for the effect of plasma lens scattering in the Galactic ISM and should be treated carefully.

In this paper, we report on the quasar 2005$+$403, seen through the Cygnus region, which exhibited a rare occurrence of AGN multiple imaging on parsec scales. This phenomenon was induced by anisotropic scattering on ionized structures in the interstellar medium. We show and analyse the source brightness distribution reconstructed from VLBA data obtained at different observing frequencies, ranging from 43~GHz down to 1.4~GHz. We investigate frequency-dependent morphology, its angular broadening and temporal changes as well as apparent spectral index distributions. Using long-term OVRO observations acquired at 15~GHz we identify and model specific flux density excursions attributed to an ESE. Through this analysis, we derive physical properties of the localised scattering screen that intersects the line of sight towards the quasar.

Throughout the paper, by the term 'core' we traditionally mean the apparent origin of AGN jet, which is typically seen as the brightest and most compact feature in VLBI images of blazars \citep[e.g.][]{Lobanov98} if not strongly affected by scattering. The spectral index $\alpha$ is defined as $S_\nu\propto\nu^\alpha$, where $S_\nu$ is the flux density measured at observing frequency $\nu$. All position angles are given in degrees east of north. We adopt the $\Lambda$CDM cosmological model with $\Omega_m=0.31$, $\Omega_\Lambda=0.69$ and $H_0=67$~km~s$^{-1}$~Mpc$^{-1}$ \citep{cosmology2020}.

\section{Searching for the refractive-dominated scattered sources}
\label{s:sampling}
Diffractive and refractive scattering of radio waves of distant source manifest themselves through different observable phenomena, but both require the presence of a highly turbulent plasma as an intermediate medium in front of a compact and bright radio source. In our previous study \citep{Koryukova2022}, we conducted an analysis of the scattering properties of the interstellar medium within the Galaxy. This analysis was based on VLBI observations of approximately 9\,000 AGN jets across frequencies ranging from 1.4 to 86~GHz. We identified a set of sources displaying extreme angular broadening at long wavelengths \citep[for the details see Section 4 in][]{Koryukova2022}. These sources were used in identifying the large-scale regions of the Galaxy containing free-electron density fluctuations, as well as associating these regions with known objects such as supernova remnants, nebulae, regions of active star formation etc. Notably, a significant concentration of scattered sources was observed within the Galactic plane ($|b|<10^\circ$), particularly near the Galactic Center and in the Cygnus constellation region \citep[Section 5, ][]{Koryukova2022}. In this study, we employ the results of our previous analysis, specifically the power-law $k$-index values derived from the frequency dependence of angular size $\theta\propto\nu^{-k}$ for a sample of 3076 all-sky AGNs. The sizes of these AGNs were measured based on  simultaneous VLBA observations at 2 and 8~GHz, with the full width at half maximum (FWHM) of the apparent jet origin (VLBI core) used as the characteristic size. The value of index $k$ effectively reflects the strength of diffractive-dominated scattering, with a value close to 1 for unscattered sources \citep{BK79,Koenigl81,KP81} and close to 2 for scattered sources \citep[e.g.][]{Rickett77}. Based on this, we considered nearly 300 AGNs with a $k$-index greater than 1.5 as potential candidates for detecting evidence of refractive-dominated scattering. We visually inspected the brightness distributions of these 300 AGNs across frequencies ranging from 1.4 to 15.4~GHz, with a particular focus on lower frequencies. AGNs displaying anisotropy in scattering were selected for further detailed analysis. Interestingly, this manual search for refractive-dominated scattered sources over a fairly large sample of AGNs yielded only a few examples, including the quasar 2005$+$403. In this paper we focus on investigating this specific object, while the results for other identified sources will be presented elsewhere.

\begin{figure}
    \centering
    \includegraphics[width=1\linewidth]{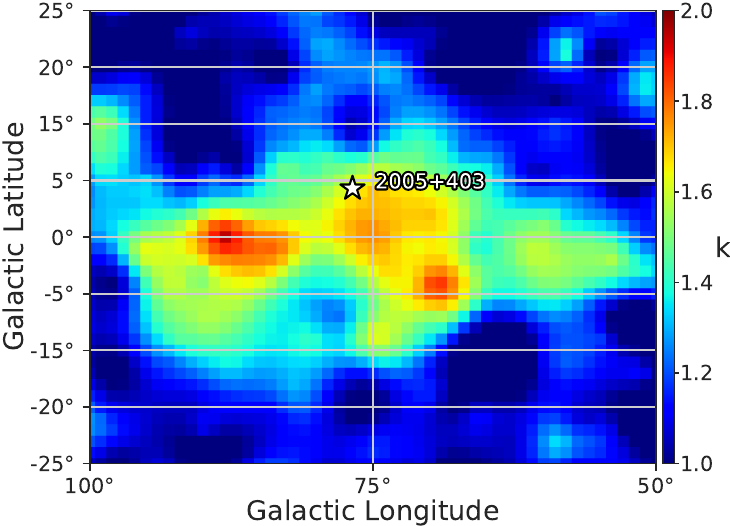}\vspace*{0cm} 
    \caption{The power-law $k$-index distribution map of the Cygnus region derived from the AGN core sizes simultaneously measured at 2 and 8~GHz ($\theta\propto\nu^{-k}$). The magnitude and colour of each pixel of the map reflect the average over $5^\circ$ area value of the $k$ index at this location. The figure is created based on the results obtained in 
    \citealt{Koryukova2022}.}
    \label{fig:k_index_vs_J2007}
\end{figure}

The line of sight towards the quasar 2005$+$403 passes through the Cygnus region, which is known for its prominent interstellar medium influence. Previous studies by e.g. \cite{Fey1989} and \cite{Gabani2006} have identified a high scattering measure along this line of sight, indicating the strong impact of the interstellar medium. The Cygnus region itself is characterized by extensive bright radio emission and encompasses objects at various stages of star formation and stellar evolution \citep{Sridharan2002, Beuther2002, Motte2007}. It also contains ultra-compact H\,II regions \citep{Downes1966, Wendker1991, Cyganowski2003}, OB stars \citep{Wright2010}, and supernova remnants \citep{Uyaniker2001}. Additionally, \cite{Pushkarev2013} have detected a rare multiple imaging event along the close line of sight, towards the quasar 2023$+$335. In \autoref{fig:k_index_vs_J2007}, we show the Cygnus region $k$-index distribution map containing about 250 AGNs which angular sizes were used for creating the map. The color-coded magnitude of each $1^\circ\times1^\circ$ pixel of the map corresponds to the average value of the $k$ index within a $5^\circ$ area around this location, weighted by a Gaussian function. Along the line of sight towards 2005$+$403, a derived $k$-value of 1.71 was obtained \citep{Koryukova2022}, indicating significant scattering of the source.

\section{Data in use}
\label{s:data}
The extragalactic radio source TXS~2005$+$403 (J2007$+$4029) is a low-spectral peaked quasar at redshift $z=1.736$ \citep{Boksenberg76} that corresponds to a scale-factor 8.55~pc~mas$^{-1}$. The analysis of parsec-scale kinematics shows superluminal motion with maximum apparent speed $9.76c$ \citep{Lister19}. The source Doppler-factor inferred from brightness temperature analysis is quite low, $\delta=3.1$ \citep{Homan21}, which is likely the reason of not been detected at high energies \citep{Lister15}. The quasar is at low Galactic latitude $b=4\fdg3$ towards the Cygnus region ($l=76\fdg8$).

\subsection{VLBA observations}
Our multi-frequency observations of 2005$+$403 were carried out quasi-simultaneously at 1.5, 1.8, 2.6 and 5.0~GHz at two epochs, on 2017 July 12 (BG246T) and 2019 February 20 (BG258G) using nine VLBA antennas (no Kitt Peak and Mauna Kea, respectively). The data were recorded with a bit rate of 2048~Mb~s$^{-1}$ and 2-bit sampling using the Mark5C acquisition system and correlated at the VLBA DiFX software correlator \citep{DifX} in Socorro, New Mexico, with 4.2~s accumulation periods. The target source had six and nine scans distributed at different hour angles that corresponded to a total integration time of 11 and 21~min, respectively for the experiments in 2017 and 2019. Additionally, we reduced publicly available archival VLBA observations of the source at higher frequencies, up to 43~GHz. We also made use of the open MOJAVE program\footnote{\url{https://www.cv.nrao.edu/MOJAVE}} data of \href{https://www.cv.nrao.edu/MOJAVE/sourcepages/2005+403.shtml}{2005+403} at 15.4~GHz. The full list of the observing experiments, data of which was used in this study, is given in \autoref{tab:data_table}.

Data reduction, including initial amplitude and phase calibration was performed with the NRAO Astronomical Image Processing System \citep[AIPS,][]{Greisen2003} following standard techniques. CLEANing \citep{Hongbom1974}, phase and amplitude self-calibration \citep{Jennison1958, Twiss1960} were performed in the Caltech \texttt{Difmap} package \citep{Shepherd_1997}. Final maps were produced by applying natural weighting of the visibility sampling function. We restored all the maps with a circular beam in order to show that the changes in brightness distributions were not due to a beam shape but produced by anisotropic scattering. The uncertainty of the obtained flux densities is assumed at the level of 10 per cent. The source structure was model-fitted in the visibility $(u,v)$ plane with \texttt{Difmap} procedure \texttt{modelfit} using a limited number of circular Gaussian components that, after being convolved with the restoring beam, adequately reproduce the constructed brightness distribution.

\begin{table}
	\caption{Observational data of the quasar 2005$+$403.}
	\centering
	\begin{tabular}{|*{3}{c|}} % four columns, alignment for each
		\hline\hline
            Epoch & Project Code & $\nu$ \\
                  &              & (GHz) \\
            (1)   &    (2)       & (3)    \\
		\hline
            1997-01-10     & BF025A & 2.27, 8.34\\
            2010-11-05     & BG196H & 1.39\\
            2012-01-02$^*$ & MOJAVE & 15.4\\
            2017-07-12     & BG246T & 1.46, 1.81, 2.56, 4.98\\
            2019-02-20     & BG258G & 1.46, 1.81, 2.56, 4.98\\
            2019-03-24     & BH222E & 1.45\\
            2019-10-15     & BP240A & 23.77\\
            2019-12-21     & BP240C & 43.17\\
		\hline
	\end{tabular}
          \begin{tablenotes}
            \item The columns are as follows: (1) observational epoch; (2) the code of the VLBA program; (3) central observing frequency.
            \item $^*$ Last epoch. All epochs at 15.4~GHz available from the MOJAVE \href{https://www.cv.nrao.edu/MOJAVE/sourcepages/2005+403.shtml}{website} were used in this work.
        \end{tablenotes}
	\label{tab:data_table}
\end{table}

\subsection{Single-dish OVRO monitoring}

OVRO 40-Meter Telescope uses off-axis dual-beam optics. The cryogenic receiver is centered at 15~GHz with 2~GHz equivalent noise bandwidth. Atmospheric and ground contributions as well as gain fluctuations are removed with the double switching technique \citep{1989ApJ...346..566R} where the source is sequentially switched between the beams. Until May 2014 the two beams were rapidly alternated using a Dicke switch, since
May 2014 a 180~degree phase switch is used in a new pseudo-correlation receiver. 

Relative calibration is obtained with a temperature-stable noise diode to compensate for gain drifts. The primary flux density calibrator is 3C~286 with an assumed value of 3.44~Jy \citep{1977A&A....61...99B}, DR21 is used as secondary calibrator source. \citet{Richards2011} describes the details of the observation and data reduction. The average sampling rate of 2005$+$403 is five flux density measurements per month.

\section{Brightness distribution and scattering}
\label{s:mapping}
When a radio wave propagates through a turbulent plasma, it encounters stochastic free electron density fluctuations $\delta n_e/n_e$. The spatial spectrum of the these density fluctuations in an ionized astrophysical plasma is commonly modeled as a power-law ${P}_{\delta n_e} (q)= {C}_{n_e}^2q^{-\beta}$, where $q_0\leq q\leq q_1$ is the wave number ranged by the corresponding inner $l_1=1/q_1$ and outer $l_0=1/q_0$ scales of turbulence, $\mathrm{C}_{n_e}$ is normalizing constant \citep{Rickett77, Cordes86, Armstrong95}, and $\beta$ is the index of power-law spectrum of electron density fluctuations. The Fourier transform of interferometric visibility function provides the observed radiation intensity distribution over the source, and the scattering angle shows $\theta_s \propto \nu^{-k}$ dependence, where $k = \beta/(\beta-2)$ for $2<\beta<4$ \citep{Lovelace1970}. The $k$-index is 2.0 ($\beta=4$) under the assumption of a thin scattering screen with a Gaussian distribution of free-electron density inhomogeneities along the line of sight \citep{Narayan1985, Cordes86}. In the case of stable and isotropic Kolmogorov turbulence in the scattering screen the $k$-index is 2.2, $\beta=11/3$ \citep{Kolmogorov41}. 

\subsection{Source structure at different frequencies}
\begin{figure*}
    \centering
    \includegraphics[width=0.45\linewidth]{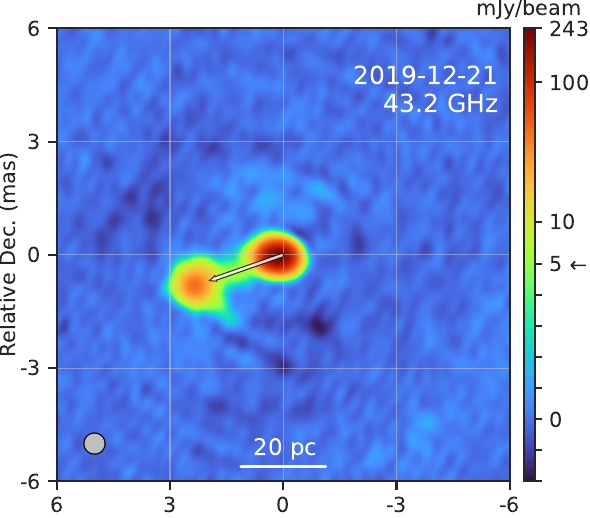}
    \hspace*{0.5cm}
    \includegraphics[width=0.44\linewidth]{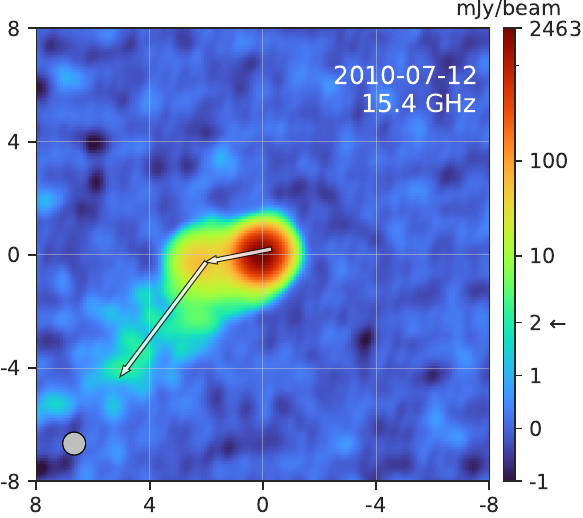}
    \includegraphics[width=0.45\linewidth]{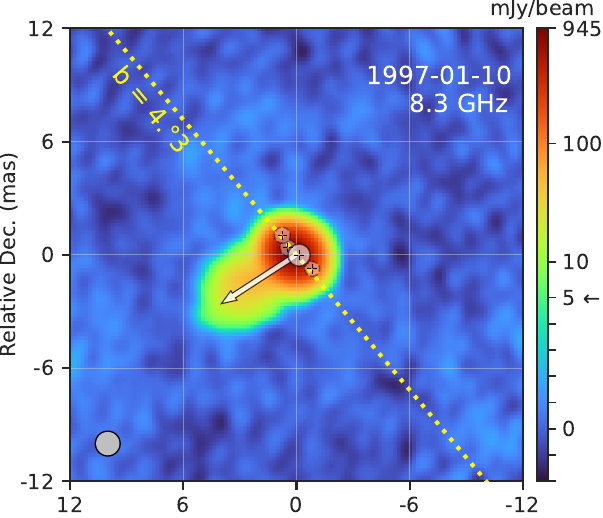}
    \hspace*{0.5cm}
    \includegraphics[width=0.435\linewidth]{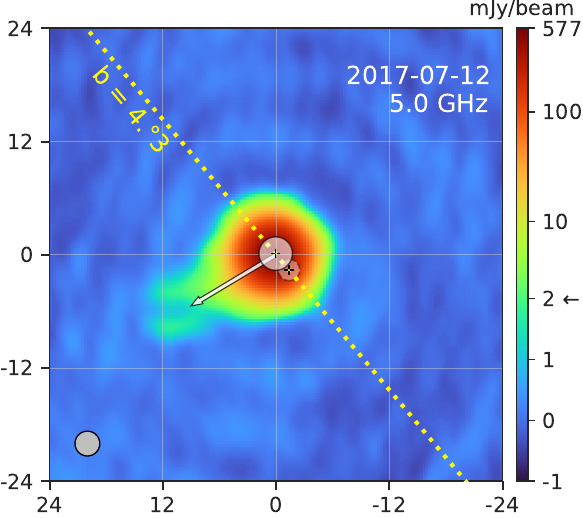}
    \includegraphics[width=0.45\linewidth]{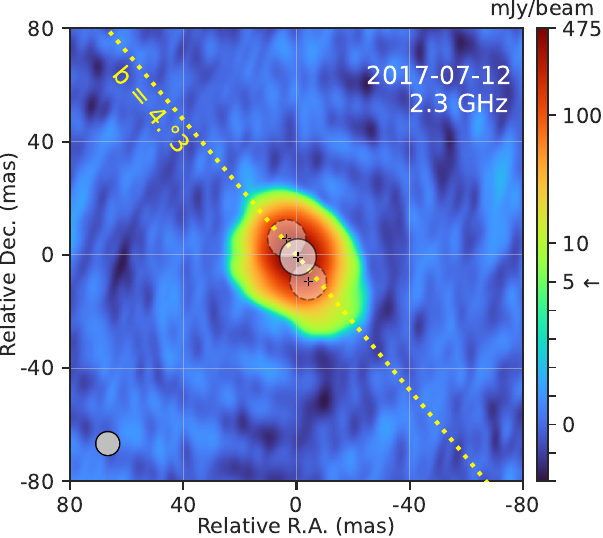}
    \hspace*{0.5cm}
    \includegraphics[width=0.44\linewidth]{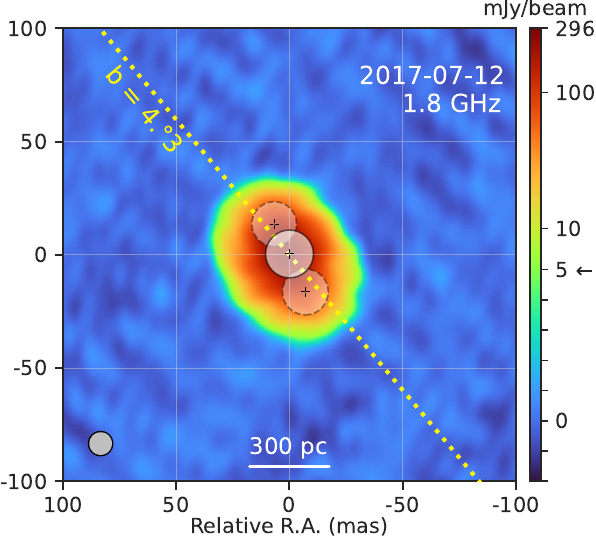}
    \caption{VLBA images of 2005$+$403. The jet direction is shown by arrow(s). Upper left: 43.2~GHz map. The jet propagates at $\mathrm{PA}=110^\circ$. Upper right: 15.4~GHz map with the inner jet propagating at $\mathrm{PA}=93^\circ$ and outer jet at $\mathrm{PA}=126^\circ$. Middle left: 8.3~GHz map with the jet developing along $\mathrm{PA}=121^\circ$. Middle right: 5.0~GHz map with the jet at $\mathrm{PA}=120^\circ$. Bottom left and right: $2.3$ and $1.8$~GHz maps, respectively. The scatter-induced patterns are extended in the direction close to the line of constant Galactic latitude ($b = 4\fdg3$) at $\mathrm{PA}=40\fdg6$ (yellow dotted line). Shaded circles represent the fitted Gaussian components of the core (brightest feature, solid line circle) and its sub-images (dotted line circle). The colorbar scale transits from linear to logarithmic at 2~mJy~beam$^{-1}$ for the 15.4 and 5.0~GHz maps, and at 5~mJy~beam$^{-1}$ for maps at other frequencies. To visualize the power of scattering at the observing frequency, the field of view for each map is matched to the size of the corresponding restoring beam, which size at FWHM is shown in the lower left corner of each image. The map parameters are listed in \autoref{tab:maps_info}.}
    \label{fig:maps}
\end{figure*}

There are two main observational effects caused by scattering. First, temporal variations in the amplitude of the signal due to the relative motion of the source, scattering medium and the observer. Second, scattering leads to the source image distortions. As the scattering strength depends on $\lambda^2$, the effects become more pronounced at lower observing frequencies. In \autoref{fig:maps}, we present total intensity VLBA images of the quasar 2005$+$403 taken at different frequencies: 43.2~GHz (BP240C, observed on 2019 December 21), 15.4~GHz (MOJAVE, 2010 July 12), and 8.3~GHz (BF025, 1997 January 10). Additionally, the quasar was simultaneously observed at lower frequencies, 5.0, 2.3 and 1.8~GHz within the BG246T session (images from BG258G and other observations are given in Appendix~\ref{appendix}). The parameters of the images are listed in \autoref{tab:maps_info}. At a relatively high observing frequency of 43.2~GHz, the source brightness distribution shows a typical one-sided core-jet AGN morphology. The source structure is modeled by four circular Gaussian components. The core component located at the apparent jet origin is the brightest and most compact feature. The other jet components progressively weaken with core separation and propagate at a position angle (PA) $110^\circ$ (\autoref{fig:maps}, upper left). 

At 15.4~GHz, the source brightness distribution shows a similar morphology (\autoref{fig:maps}, upper right), the best-fit model of which contains five Gaussian components. The brightest feature is one of the inner jet components, not the core. The jet of 2005$+$403 develops in $\mathrm{PA}=93^\circ$ within first 2~mas, then changes its direction to South-East, with $\mathrm{PA}=126^\circ$. As noted by \cite{Gabani2006}, paths of the innermost jet components is curved and their motion is not ballistic, suggesting a spatially curved (helical) jet. Beyond 2~mas core-separation, however, the components seem to move on linear (ballistic) trajectories \citep{Lister19}.

At 8.3~GHz and 5.0~GHz (\autoref{fig:maps}, middle left and right) we begin to observe clear indications of scattering, although the jet is still revealed at $\mathrm{PA}=123^\circ$ and $\mathrm{PA}=120^\circ$, respectively. In addition to the core and jet features, we detected additional components located at PA of $57^\circ$, $40^\circ$ and $-137^\circ$ relative to the core. At 5.0~GHz we detect only one additional component located at PA of $-141^\circ$ relative to the core. We can also note a strong influence of isotropic diffractive-dominated scattering, that broadens the observed brightness distribution of the source in all directions.

\begin{table*}
	\caption{Map parameters from \autoref{fig:maps}.}
	\centering
	\begin{tabular}{|*{9}{c|}} 
		\hline\hline
            Epoch & $\nu$ & $I_\mathrm{peak}$ & $S_\mathrm{tot}$& rms              & Beam   & Pixel size \\
                  & (GHz) & (mJy beam$^{-1}$) & (Jy)            & (mJy beam$^{-1}$)&  (mas) & (mas)      \\
            (1)   & (2)   & (3)               & (4)             & (5)              & (6)    & (7)        \\
		\hline
            2019-12-21 & 43.2  & 243.1   & 0.46 & 0.27 & 0.53 & 0.02\\       
            2010-07-12 & 15.4  & 2463.1  & 3.85 & 0.24 & 0.75 & 0.1\\
            2010-07-12 & 8.3   & 944.9   & 1.97 & 0.45 & 1.23 & 0.2\\
            2017-07-12 & 5.0   & 577.4   & 1.85 & 0.21 & 2.45 & 0.3\\
            2017-07-12 & 2.3   & 475.5   & 2.21 & 0.53 & 7.95 & 0.5\\
            2017-07-12 & 1.8   & 296.3   & 2.03 & 0.37 & 9.98 & 0.8\\
		\hline
	\end{tabular}
          \begin{tablenotes}
            \item The columns are as follows: (1) observational epoch; (2) central observing frequency; (3) intensity peak of the map; (4) total flux density of the map; (5) the root mean square level of the residuals of the final map; (6) FWHM size of the restoring circular beam; (7) pixel size. 
            \end{tablenotes}
	\label{tab:maps_info}
\end{table*}

At low frequencies (1.4--2.3~GHz) the observed morphology of the quasar becomes highly atypical (\autoref{fig:maps}, bottom panels). The jet emission is no longer visible. Instead, the restored brightness distribution appears to be elongated nearly along the line of constant Galactic latitude $b = 4\fdg3$ of the source (\autoref{fig:maps}, yellow dotted line in the $\mathrm{PA} = 40\fdg6$), suggesting that the emission could be induced by anisotropic refractive-dominated scattering. According to theoretical predictions \citep{Clegg_1998}, the refraction of radio waves in the interstellar medium can create a hierarchy of (sub-)images (up to three including the primary image for a lens with a Gaussian density profile) displaced in the projected direction from the background source to the scattering screen. Structure model fitting shows that the secondary images of the compact core feature are several times weaker in flux density but have comparable angular sizes (\autoref{tab:modelfits}). In cases where two sub-images of the core are detected, they are typically located at comparable separations from it but in opposite directions (e.g., \autoref{fig:maps}, bottom panel). This could be caused by multi-component or clumpy structure of the scattering screen \citep{Pushkarev2013} creating an edge effect on both sides of the core component. 

In \autoref{tab:modelfits}, we present the model fitting results of the data from \autoref{fig:maps}. The maps and corresponding model fit results from other data sets are given in \autoref{afig:maps} and \autoref{atab:modelfits}, respectively. The errors of the obtained modelfit parameters were estimated from the image plane using the following analytical approximations \citep[][and references therein]{Schinzel2012}

\begin{gather}
\sigma_\mathrm{\theta} = \frac{\sigma_\mathrm{peak}}{I_\mathrm{peak}}\left(\theta^2+b_\mathrm{maj}b_\mathrm{min}\right)^{1/2}
	\label{eq:eq1},\\
\sigma_\mathrm{r} = \frac{1}{2}\sigma_\mathrm{\theta} 
	\label{eq:eq2},\\
\sigma_\mathrm{peak} = \sigma_\mathrm{rms}\left(1+\frac{I_\mathrm{peak}}{\sigma_\mathrm{rms}}\right)^{1/2}
	\label{eq:eq3},\\
\sigma_\mathrm{tot} = \sigma_\mathrm{peak}\,\frac{S_\mathrm{tot}}{I_\mathrm{peak}}
	\label{eq:eq4},\\
\sigma_\mathrm{\varphi} = \mathrm{atan}\biggl(\frac{\sigma_{r}}{r}\biggr)
	\label{eq:eq5},\\
\theta = \mathrm{max}(\mathrm{component\ size;\ resolution\ limit})
	\label{eq:eq6},
\end{gather} 

\noindent
where the resolution limit \citep{Kovalev05}
\begin{equation}
\theta_\mathrm{lim} = 2\,b_\varphi\left[\frac{\ln2}{\pi}\ln\left(\frac{\mathrm{SNR}}{\mathrm{SNR}-1}\right)\right]^{1/2}, \end{equation}

\noindent
where $\mathrm{SNR}=I_\mathrm{peak}/\sigma_\mathrm{rms}$ is the signal-to-noise ratio of a component, $b_\varphi$~-- the half-power beam size along a position angle $\varphi$ of a component, $\theta$~-- component size, $r$~-- its separation from the core, $S_\mathrm{tot}$~-- total flux density of a component in Jy, $I_\mathrm{peak}$~-- peak intensity of a component in Jy~beam$^{-1}$, $\sigma_\mathrm{rms}$~-- post-fit rms in Jy~beam$^{-1}$ around component position in the residual image.  

\begin{figure}
    \centering
    \includegraphics[width=\linewidth]{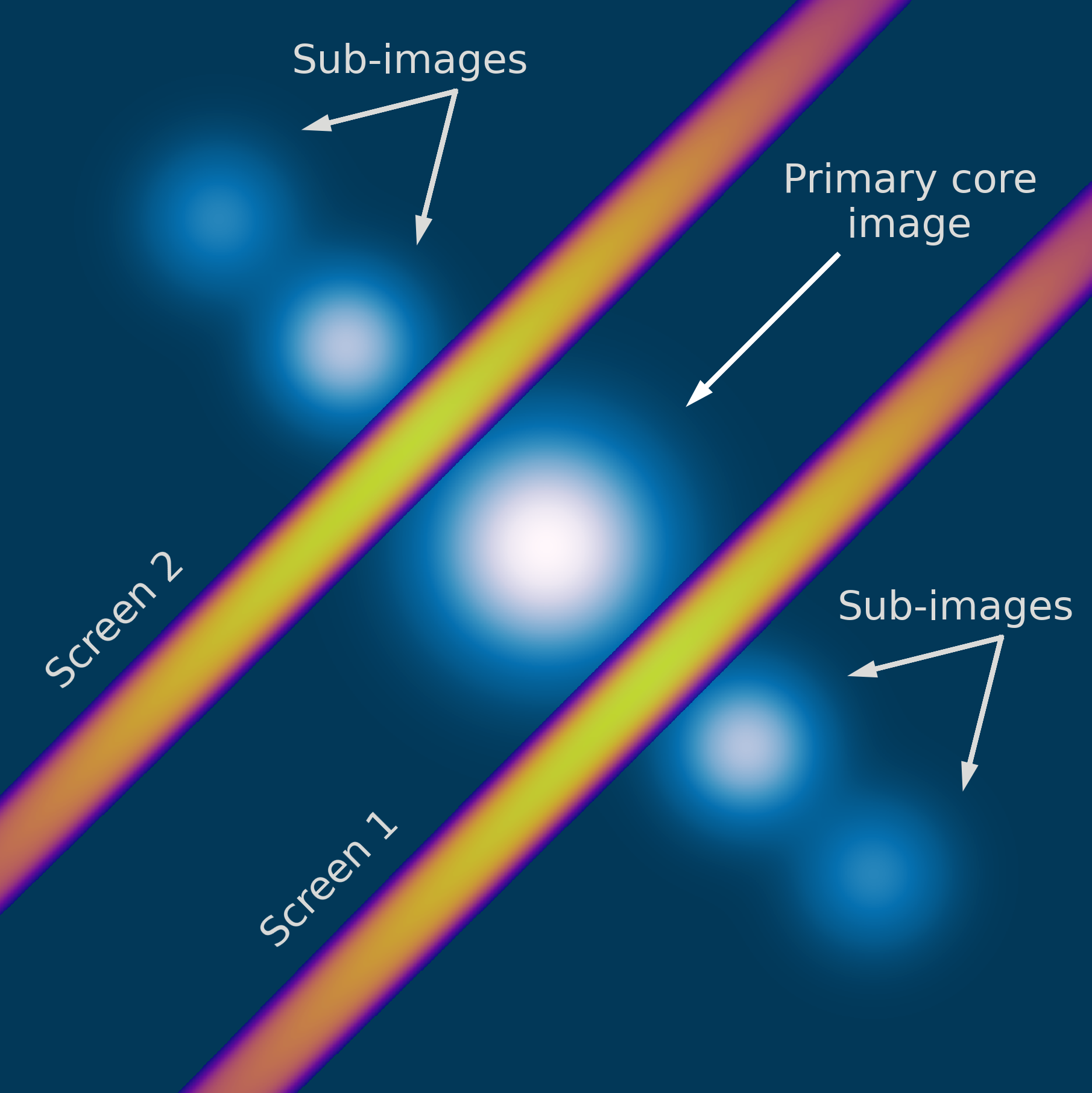}
    \caption{Schematic drawing that shows the formation of a hierarchy of refraction-induced sub-images on the opposite sides of the background source in the case of a flat geometry of the scattering screen with a transverse Gaussian profile of electron density.}
    \label{fig:scattering_screen_schem}
\end{figure}

Notably, given the presence of refractive-dominated scattering, it forms the secondary images at different frequencies and epochs along the preferred direction of the constant Galactic latitude line. This implies that (i) the screen material is moving parallel to the Galactic plane, (ii) it likely has a flat geometry, e.g. shock fronts, (iii) there are multiple scattering screens along the line of sight, at least two on opposite sides of the background source. The latter might be the reason why at some epochs the VLBI core position is uncertain. We schematically depicted a hierarchy of multiple source images, when the edge effect is created by two flat geometry screens with a transverse density profile close to Gaussian on opposite sides of the source (\autoref{fig:scattering_screen_schem}), forming a pattern of one prime and four secondary images of the lensed object. Alternatively, this might occur if both refractive and diffractive effects are acting together, comparably influencing the source emission. For example, this is evident in the 2.3~GHz data at epoch 1997 January 10 (\autoref{afig:maps}, middle left). If the scattering screen has a curved shape/boundary, secondary images can be induced in various PAs, forming an asymmetrical observed structure of the source. In Appendix~\ref{appendix} we present the map and model of the quasar on November 5, 2010 obtained at 1.4~GHz, where one of the sub-images is located at $\mathrm{PA}=-5\fdg7$ (\autoref{afig:maps}, continued, upper left). We hypothesize that any non-symmetrical alignment of sub-images relative to the core may be caused by the curvature of the scattering screen, characterized by a transverse Gaussian free-electron density distribution. The secondary images at 1.5 and 1.8~GHz on February 20, 2019 (\autoref{afig:maps}, bottom) are also potential cases of this scenario. However, these models are complicated for such analysis due to the uncertainty of VLBI core position.

\begin{table*}
	\caption{The results of brightness distribution model fitting using circular Gaussian components for the data illustrated in \autoref{fig:maps}.}
	\centering
	\begin{tabular}{|*{7}{c|}} 
		\hline
		\hline
		Epoch          & $\nu$ & Comp. & $S$           &  $r$           & $\phi$          & $\mathrm{\theta}$\\
                       & (GHz) &       & (Jy)          & (mas)          & ($^\circ$)      & (mas) \\
            (1)        & (2)   & (3)   & (4)           & (5)            & (6)             & (7) \\
		\hline
            2019-12-21 & 43.2  & core  & $0.19\pm0.02$ & 0.00           & \ldots          & $0.12\pm0.05$  \\ 
                       &       & jet   & $0.14\pm0.02$ & $0.34\pm0.03$  & $90.9\pm5.2$    & $0.19\pm0.06$  \\ 
                       &       & jet   & $0.08\pm0.01$ & $2.50\pm0.05$  & $108.2\pm1.2$   & $0.48\pm0.10$  \\ 
                       &       & jet   & $0.05\pm0.03$ & $1.63\pm0.50$   & $103.7\pm17.0$  & $1.52\pm1.00$  \\ 
            \hline
            2010-07-12 & 15.4  & core  & $0.90\pm0.06$ & 0.00           & \ldots          & $0.40\pm0.03$  \\
                       &       & jet   & $1.36\pm0.06$ & $0.40\pm0.02$  & $138.3\pm2.6$   & $0.32\pm0.04$  \\ 
                       &       & jet   & $0.93\pm0.05$ & $0.36\pm0.02$  & $96.4\pm3.2$    & $0.26\pm0.04$  \\ 
                       &       & jet   & $0.47\pm0.06$ & $0.53\pm0.07$  & $117.8\pm7.9$   & $0.95\pm0.15$  \\ 
                       &       & jet   & $0.19\pm0.05$ & $2.31\pm0.18$  & $101.4\pm4.5$   & $1.21\pm0.36$  \\
            \hline
            1997-01-10 & 8.3   & core  & $1.45\pm0.08$ & 0.00           & \ldots          & $1.18\pm0.10$   \\ 
                       &       & sub-c & $0.40\pm0.04$  & $0.71\pm0.07$  & $56.6\pm5.8$    & $0.84\pm0.14$  \\ 
                       &       & jet   & $0.24\pm0.07$ & $2.78\pm0.37$  & $121.1\pm7.7$   & $2.42\pm0.75$  \\ 
                       &       & sub-c & $0.06\pm0.02$ & $1.00\pm0.18$   & $-136.8\pm10.2$ & $0.79\pm0.36$  \\ 
                       &       & sub-c & $0.05\pm0.01$ & $1.38\pm0.19$  & $40.1\pm7.7$    & $0.84\pm0.37$  \\ 
            \hline
            2017-07-12 & 5.0   & core  & $1.69\pm0.13$ & 0.00           &  \ldots         & $3.54\pm0.33$  \\ 
                       &       & sub-c & $0.10\pm0.02$ & $2.24\pm0.39$  & $-141.0\pm9.9$  & $2.27\pm0.78$  \\ 
                       &       & jet   & $0.05\pm0.03$ & $4.95\pm1.22$  & $130.2\pm13.9$  & $4.25\pm2.45$  \\         
            \hline
            2017-07-12 & 2.3   & core  & $1.38\pm0.14$ & 0.00           & \ldots          & $12.87\pm1.48$ \\ 
                       &       & sub-c & $0.55\pm0.09$ & $7.60\pm1.28$  & $31.4\pm9.6$    & $13.42\pm2.57$ \\ 
                       &       & sub-c & $0.28\pm0.06$ & $9.38\pm1.63$  & $-157.4\pm9.8$  & $12.84\pm3.25$ \\	
            \hline
            2017-07-12 & 1.8   & core  & $1.44\pm0.17$ & 0.00           & \ldots          & $21.08\pm2.70$ \\ 
                       &       & sub-c & $0.29\pm0.08$ & $18.16\pm3.05$ & $-157.4\pm9.5$  & $20.19\pm6.09$ \\ 
                       &       & sub-c & $0.30\pm0.07$ & $14.84\pm2.48$ & $27.05\pm9.5$   & $19.72\pm4.96$ \\ 
            \hline
	\end{tabular}
         \begin{tablenotes}
            \item The columns are as follows: (1) epoch of observations; (2) central observing frequency; (3) type of a component, where 'sub-c' means 'sub-component' of the core; (4) measured flux density of a component; (5) radial distance of a component relative to the core; (6) position angle of a component with respect to the core; (7) FWHM of the measured size of a component.
        \end{tablenotes}
	\label{tab:modelfits}
\end{table*}

\subsection{Special considerations for low-frequency imaging}
If a source at a given frequency is not subject to angular broadening due to scattering, a set of CLEAN boxes can be uploaded from the very beginning of imaging if known a priori. Alternatively, it can be formed gradually as the source model becomes more complete by adding areas with progressively lower intensity. Either of these approaches works equally well. However, it is not so if the observed source morphology is affected by heavy scattering, especially at low ($\lesssim5.0$~GHz) frequencies. In such cases, in order to properly reconstruct the source brightness distribution, it is crucial to use a complete set of CLEAN boxes right from the very first steps of the imaging process. Thus, it requires several full runs of the mapping procedure to iteratively create an array of CLEAN boxes, making it more complete with each subsequent run. This method ensures to (i) preserve the amplitude scale and (ii) restore the complete brightness distribution of the source with the lowest noise level. In particular, employing this approach allowed us to reach a high dynamic range of the 5.0~GHz image and reveal a faint jet (\autoref{fig:maps}, middle right).

\subsection{Angular broadening of the core}
\label{s:k_scat_fit}
\begin{figure*}
    \centering
    \includegraphics[width=0.5\linewidth]{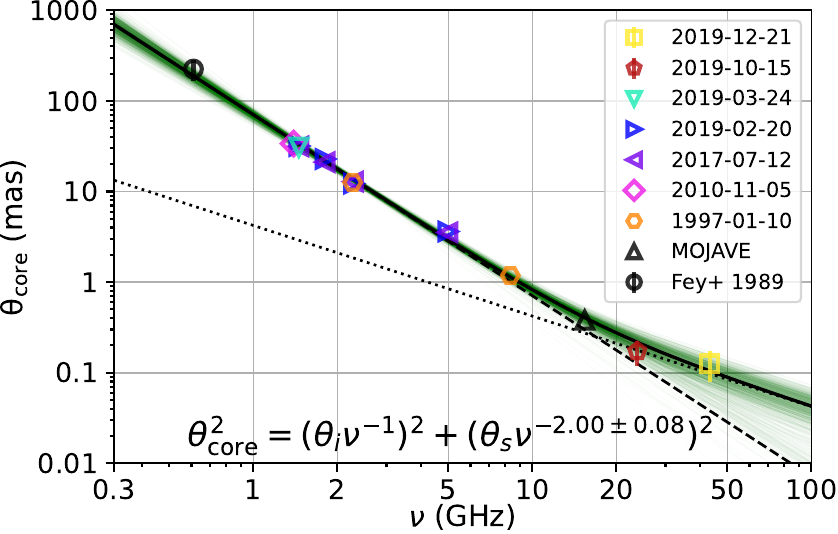}
    \includegraphics[width=0.49\linewidth]{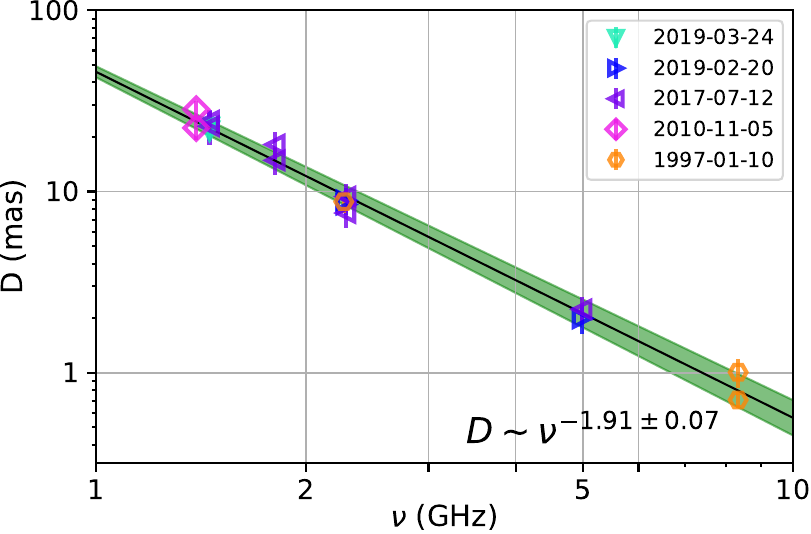} 
    \caption{Left: Frequency dependence of the observed core size. The solid line shows the fit of Eq.~\ref{eq:observed_size}, while the dashed and dotted lines represent the inferred scattered and intrinsic core size, respectively. Solid green lines indicate Monte Carlo fits. There is no core size error bar for the measurements where it is smaller than the marker size. Right: Distance between the core component and its secondary image(s) as a function of observing frequency. Different shapes and colors of markers denote different observing epochs. Green shaded area represents $1\sigma$ confidence region of the fit.}
    \label{fig:size_and_D_vs_freq}
\end{figure*}

The observed brightness distribution of a source can be described as a convolution of its intrinsic structure with the scattering function. To analyze the frequency dependence of the angular size of the AGN VLBI core, we use the equation: 
\begin{gather}
    \theta^2 = (\theta_\mathrm{i}\nu^{-1})^2 + (\theta_\mathrm{s}\nu^{-k})^2 
	\label{eq:observed_size},
\end{gather} 

\noindent
where $\theta_\mathrm{i}$ and $\theta_\mathrm{s}$ are the intrinsic and scattered angular size of a source at 1~GHz and $k$ is the scattering index. We fitted Eq.~\ref{eq:observed_size} using the VLBI core sizes of the source measured at different frequencies. At 15~GHz we used data at 31 available epochs, taking the average value of the measured size. Additionally, we  included the angular size $225\pm58$~mas of the quasar measured at 0.61~GHz by \cite{Fey1989} who fitted the source structure with a single circular Gaussian. The result of fitting Eq.~\ref{eq:observed_size} is shown in \autoref{fig:size_and_D_vs_freq} (left). The derived scattering index $k = 2.00\pm0.08$, which corresponds to the theoretical prediction for the diffractive scattering of radio waves passing through a lens with a Gaussian distribution of free-electron density \citep{Narayan1985, Cordes86}. The intrinsic and scattered angular sizes at 1~GHz are $4.2\pm1.5$ and $70.1\pm5.7$~mas, respectively. In our previous work \citep[Section~6.1,][]{Koryukova2022}, the measured core sizes of 130 AGNs (including 2005$+$403) seen through the Galactic plane ($|b|<10^\circ$) at 2, 5 and 8~GHz were used to estimate the characteristic $k$-value for the scattered sources, resulting in $k = 2.01\pm0.13$. This value supports the Gaussian screen model too but we can not exclude the Kolmogorov spectrum as well with a canonical $k=2.2$, which corresponds to $\beta=11/3$ \citep{Kolmogorov41}. 

Indeed, we have established that the derived $k$-index is dependent on the frequency range of the data. As shown in \autoref{tab:k_diff_freq_range}, when limiting the frequency range starting at 8.3~GHz and below, the progressively higher $k$-index values are obtained. Lower observing frequencies, being more sensitive to scattering, probe larger spatial inhomogeneities where the spectrum reaches the Kolmogorov's $k = 2.2$. This suggests that the choice of frequency range may affect the interpretation of the scattering index in favor of one or another turbulence model.

\begin{table}
	\caption{The power-law $k$-index fitting results according to Eq.~\ref{eq:observed_size} depending on different observing frequency ranges.}
	\centering
	\begin{tabular}{|*{3}{c|}}
		\hline
		\hline
            $\nu_\mathrm{upper}$ & $k$-index & Error \\
             (GHz)      &                     &       \\
              (1)       &      (2)            & (3)   \\
		\hline
              43.2      &      2.00           & 0.08 \\
              23.8      &      1.93           & 0.08 \\
              15.4      &      1.95           & 0.10 \\
               8.3      &      2.09           & 0.21 \\
               5.0$^*$  &      1.96           & 0.11 \\              
               2.3$^*$  &      2.16           & 0.20 \\
               1.8$^*$  &      2.17           & 0.29 \\
               1.5$^*$  &      2.27           & 0.35 \\
		\hline
	\end{tabular}
          \begin{tablenotes}
            \item The columns are as follows: (1) upper limit of frequency range started at 0.61~GHz; (2) the result of $k$-index fitting; (3) the error of $k$-index value.
            \item $^*$ Low frequency data were fitted according to $\theta_\mathrm{core} = \theta_s\nu^{-k}$.
        \end{tablenotes}
	\label{tab:k_diff_freq_range}
\end{table}

The results obtained for the scattering index $k$ in our study are in good agreement with findings from previous investigations by e.g., \cite{Fey1989}, \cite{Desai2001} and \cite{Gabani2006} for the same source. In their studies, they also analyzed the source structure using VLBI observations but fitted it with a single circular or elliptical Gaussian component in the visibility plane. The scattering index (they refer to it as 'size spectral index') value was obtained by analysing the measured total angular size of the quasar 2005$+$403 as a function of the observing frequency. \cite{Fey1989} inferred the sizes from VLBI observations at 0.6, 1.7 and 5.0~GHz and obtained $k = 1.94 \pm 0.15$. \cite{Desai2001} measured the source size at 1.7, 2.3 and 5.0~GHz, and found that $k=2.0$. Similarly, \cite{Gabani2006} used VLBI archival data covering a period from 1992 to 2003 at frequencies ranging from 1.6 to 43~GHz, and they derived $k = 1.90\pm0.05$. To make a more accurate comparison with these findings, we modeled the source structure with a single circular Gaussian at frequency of 5.0~GHz and below, and derived $k=2.02\pm0.09$.

We should particularly emphasise that the quasar 2005$+$403 has a potential to manifest a noticeably large angular size when observed with the low-frequency instruments, such as LOFAR (Low-Frequency Array) and SKA (Square Kilometre Array). LOFAR operates in a frequency range from 0.07 up to 0.2~GHz. Based on the results of this section, the expected observed angular size of 2005$+$403 is in the range from 1.7 to 14.2~arcsec. The lower frequency domain of SKA is expected to reach 50~MHz. In this case, the angular size of the source can be quite impressive, around 27.8 arcseconds, roughly 100 times larger than its intrinsic size.

\subsection{Core sub-image offsets}
The secondary images of the VLBI core are detected at frequencies ranging from 1.4 through 8.3~GHz. They are all positioned nearly along the line of constant Galactic latitude along the PA of $40^\circ$ and/or $-140^\circ$ directly indicating the orbital motion of the plasma lenses in their Galactic rotation. However, for the secondary images to be clearly separated from the core component and detectable by VLBI observations, the refractive strength of the scattering screen must be sufficiently high. The greater the free-electron density in the lens, the higher the refraction angle of radio waves will be, leading to larger separations between the core and its secondary images. The characteristics of the medium responsible for strong interstellar refraction are not yet fully understood. It is uncertain whether the refraction results from a localized large-amplitude electron density enhancement located between the observer and the source or whether it is the consequence of an extended intervening turbulent medium with numerous free-electron density fluctuations distributed along the line of sight. To probe the nature of this observed phenomenon further, we examined the frequency dependence of the separation $D_{\nu}$ between the VLBI core and its sub-image(s). By fitting this distance with a function proportional to $\nu^{-k_D}$, a power-law index $k_D$ of $1.91\pm0.07$ was obtained (\autoref{fig:size_and_D_vs_freq}, right).

We excluded from this analysis the data on February 20, 2019 at 1.5 and 1.8~GHz because (i) at this epoch the identification of the core component is complicated due to strong scattering (marked with '*' symbol in \autoref{atab:modelfits}), (ii) adding these measurements to the fit changes the resulting k value significantly. This scaling of the separation between the core and its sub-images being close to $\nu^{-2}$ provides compelling observational evidence that the extended morphology observed nearly along the line of constant Galactic latitude is formed by scattering in the ISM and is not intrinsic to the source.

\subsection{Temporal changes in brightness distribution}

\begin{figure*}
    \centering
    \includegraphics[width=0.34\linewidth]{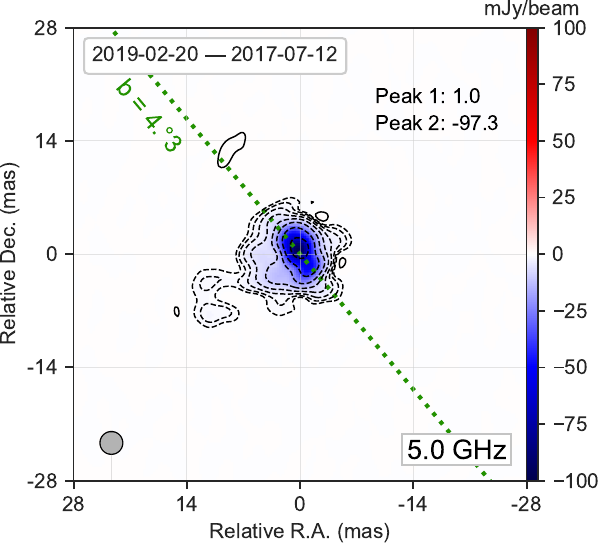}
    \includegraphics[width=0.321\linewidth]{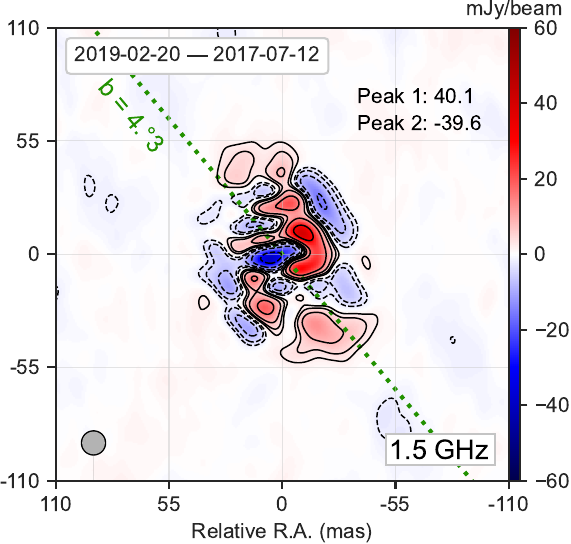}
    \includegraphics[width=0.321\linewidth]{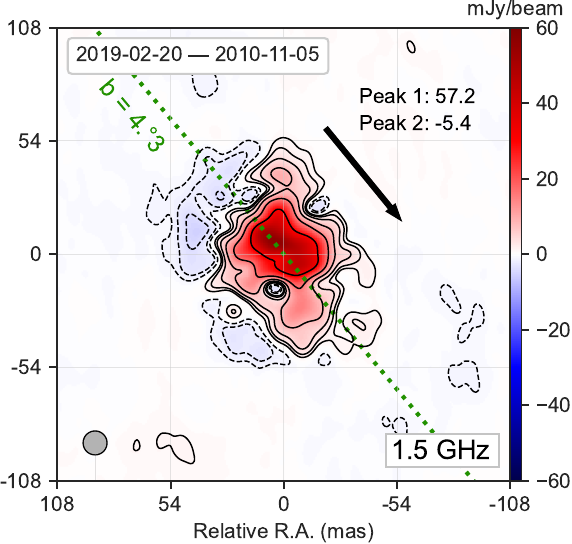}
    \caption{Difference maps in total intensity at 5.0~GHz (left) and 1.5~GHz at short (1.61~yr, middle) and long-term (8.3~yr, right) time scale. The contours are plotted at increasing power of 2, starting from the corresponding 4 times rms level. The green dotted line is line of constant Galactic latitude ($b = 4\fdg3$) at $\mathrm{PA}=40\fdg6$. The arrow shows the direction of motion of the scattering screen that crosses the line of sight towards the source, provided that scattering occurs in the Cygnus region.}
    \label{fig:diff_map}
\end{figure*} 

As the scattering screen is drifting across the line of sight to the background source, its properties are continuously changing with time, accordingly affecting the refraction-induced patterns of brightness distribution elongated along the line of constant Galactic latitude. Thus, the differential maps constructed at different frequencies and time scales can be used as a diagnostic tool for probing screen characteristics. For these purposes, we used data from the BG246T and BG258G experiments separated by 1.61~yr as they have identical frequency setups (1.5, 1.8, 2.3, and 5.0~GHz). Additionally, we made use of data at 1.4~GHz from BG196H to probe larger time scales, 6.7 and 8.3~yr.

In order to correctly interpret induced morphological changes in the source brightness distribution due to scattering, we constructed maps at the same frequency with an average circular restoring beam. These maps were superimposed by the phase centre, since the exact position of the core component is often complicated, especially at low frequencies. \autoref{fig:diff_map} presents the differential brightness distributions at 5.0 and 1.5~GHz (at two different time scale, minimum of 1.61~yr and maximum of 8.3~yr). The differential image at 5.0~GHz (\autoref{fig:diff_map}, left) constructed between the closest available epochs, 2019 February 20 and 2017 July 12, shows the structure stretched out along the $b=const$ line with only negative values. This implies that the screen, which formed the refraction-induced emission at earlier epoch either moved away completely or became significantly weaker. At 1.5~GHz (\autoref{fig:diff_map}, middle), the difference map taken between the same pair of relatively close epochs, only 1.61~yr apart, shows a total intensity redistribution at a low level, with interlaced negative and positive emission regions. This suggests that the screen properties have not essentially changed over this period of time. However, when a wider time window of 8.3~yr is considered, the difference image at 1.5~GHz (\autoref{fig:diff_map}, right) reveals clear evidence for a change in the scattering screen position relative to the core feature. We note that the screen drifts in a PA of about $-140^\circ$ assuming it is located in the Cygnus region. If the screen is situated at larger distances, e.g. in the Perseus Arm or the Outer Arm, its motion in the projection on the sky would be in the opposite direction.

\begin{figure*}
    \centering
    \includegraphics[width=0.39\linewidth]{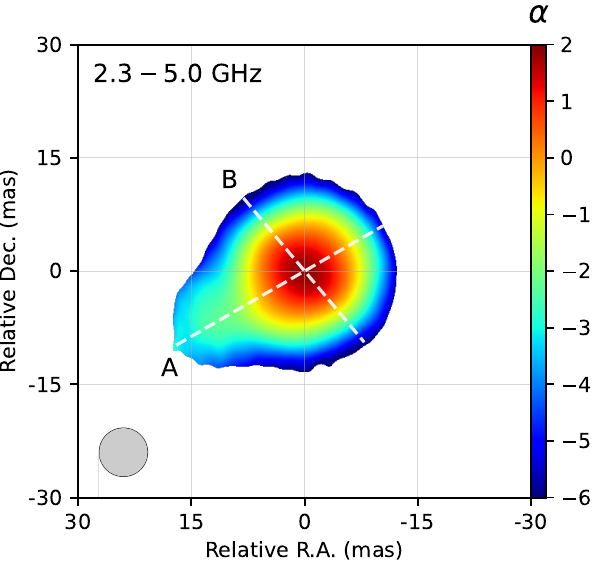}
    \hspace*{0.1cm}
    \includegraphics[width=0.39\linewidth]{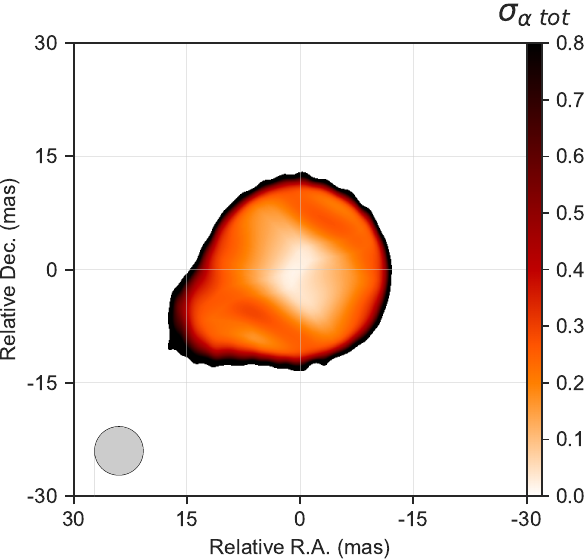}\vspace*{0.5cm}
    \includegraphics[width=0.36\linewidth]{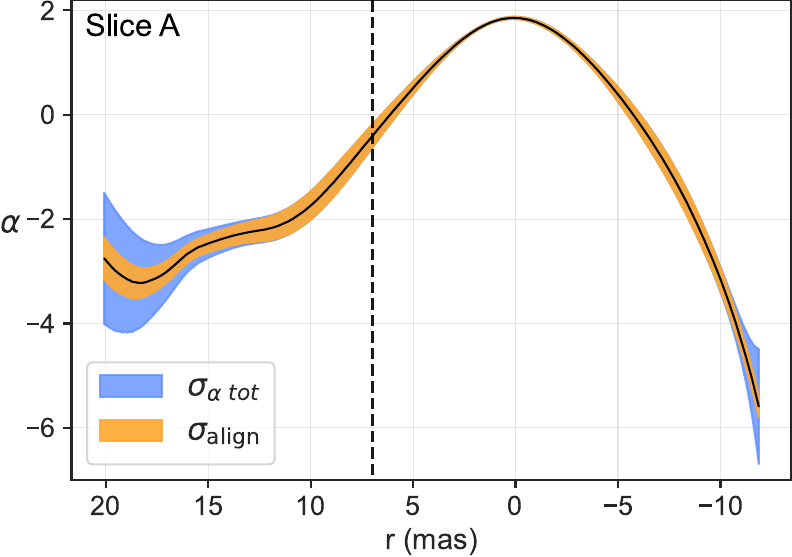}
    \hspace*{0.8cm}
    \includegraphics[width=0.36\linewidth]{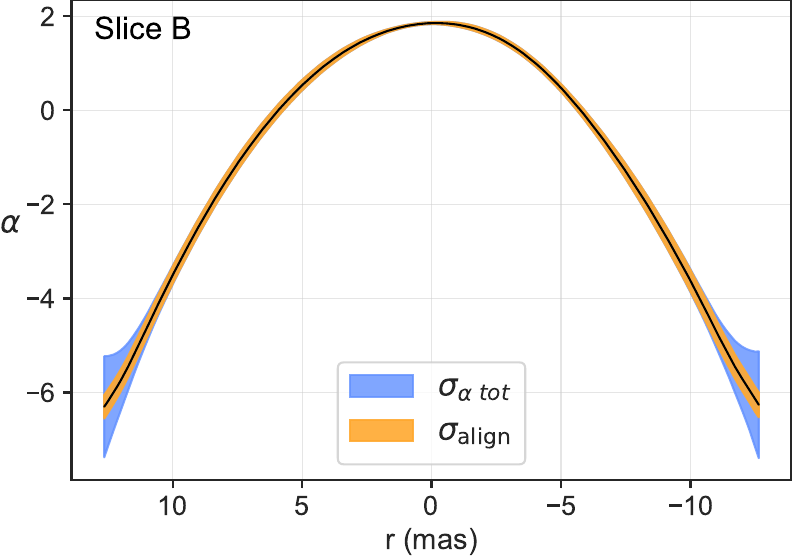}
    \caption{Upper panel: (Left) The spectral index distribution map calculated using the simultaneous observational data at 2.3 and 5.0~GHz at epoch 2017 July 12. We aligned the maps by the phase center applying no shift; (Right) The spectral index total error distribution map calculated with the contribution of systematic, random and align error. The grey circle in the bottom left corner is the FWHM of average restoring beam. Dashed lines depict profiles in the direction of (A) jet/counter-jet at $\mathrm{PA}=120^\circ$, and (B) in the direction of the line of constant Galactic latitude $b = 4\fdg3$ at $\mathrm{PA}=40\fdg6$. Bottom panel: Spectral index profiles in (A) and (B) directions. The shaded area represents the total error of $\alpha_\mathrm{tot}$ (blue) and map aligning error $\sigma_\mathrm{align}$ (orange). The dashed line marks the approximate core separation, beyond which the spectral index distribution is not significantly influenced by scattering.
}
    \label{fig:spectral_index}
\end{figure*}

\section{Spectral properties}
\subsection{Spectral index distribution maps}
\label{s:spectral_index}

Using the simultaneously obtained data at frequencies 1.4, 1.8, 2.3, and 5.0~GHz and corresponding brightness distributions of the source from the BG246T and BG258G experiments, we constructed spectral index maps at different pairs of frequencies and analysed them. We used the image-plane data extracting information from the maps restored with the average circular beam. The circular beam averaged over the original beam sizes at the selected pair of frequencies. In the case of a strongly scattered source, we can not align maps by an optically thin region, as it is usually done for AGN jets, since the position of this region is no longer achromatic and thus uncertain due to heavy scattering, especially at frequencies lower than 5.0~GHz. Therefore, we aligned the maps by the phase center applying no shift. We created two-frequency spectral index distribution maps by calculating its value in each pixel (pixel size is 0.1~mas). We blank the pixels where total intensity is less than 3\,rms at any frequency. The total error of the spectral index at each pixel was estimated as \(\sigma_\mathrm{\alpha\,tot} = \sqrt{\sigma_\mathrm{\alpha}^2+\sigma_\mathrm{align}^2}\), where $\sigma_\mathrm{align}$ is the error of map alignment and $\sigma_\mathrm{\alpha}$ is the spectral index error in each pixel calculated as
\begin{gather}
    \sigma_\mathrm{\alpha} = \frac{1}{|\mathrm{ln}(\nu_1/\nu_2)|}\sqrt{\left(\frac{\sigma_{I_1}}{I_1}\right)^2+\left(\frac{\sigma_{I_2}}{I_2}\right)^2}
	%\label{eq:alpha_err},
\end{gather} 

\noindent
where $\sigma_I = \mathcal{E} I_{i,j} + \sigma_\mathrm{rms}$ is the total intensity error in each pixel of the map (i,\,j) that considered to be the sum of systematic amplitude error and thermal random noise obtained as rms over the total intensity map excluding the regions of source emission. The systematic amplitude error is assumed to be at the level of $\mathcal{E}=10\%$ of total intensity. As the image registration accuracy is unknown, we assessed $\sigma_\mathrm{align}$ in each pixel as the mean of absolute difference in $\alpha$ derived from the intensity maps at different frequencies shifted by 1~mas in eight specific directions: along the jet ($\mathrm{PA}=120^\circ$), counter-jet, two normals to the jet direction, and four directions along right ascension and declination.

All constructed spectral index distributions show similar features, therefore we present and discuss only one of them. In \autoref{fig:spectral_index} (top left), we show the spectral index map calculated using the data simultaneously obtained on July 12, 2017 at 2.3 and 5.0~GHz within the BG246T experiment. Typically, spectral index distributions for parsec-scale AGN jets show a flatter spectrum in the core region, while the optically thin emission of the jet has a steeper spectrum, with a mean value of $+0.22\pm0.03$ and $-1.04\pm0.03$, respectively \citep[e.g.,][]{PushkarevKovalev2012, HovattaAller2014}. Scattering effects being strongly frequency dependent ($\propto\nu^{-2}$) affect the emission at 2.3~GHz much more heavily than that at 5.0~GHz. This leads to a number of atypical features on the corresponding distribution of spectral index. First, $\alpha$ values span a wide range, from about $-6$ at the brightness distribution edges to $+2$ in the center area, not upstream the core at lower frequency as we typically observe for the unscattered sources. Second, the most extreme negative values are revealed at the edges towards secondary images of the core, roughly orthogonal to the jet direction. Third, the spectral index distribution in the core area is quite symmetric. In \autoref{fig:spectral_index} (bottom panel), we depict profiles of the spectral index in directions of the line of constant Galactic latitude and in the direction of the jet/counter-jet. Slice B (\autoref{fig:spectral_index}, bottom right) clearly shows the symmetry along the line of constant Galactic latitude, while Slice A (\autoref{fig:spectral_index}, bottom left) is asymmetric due to the observed jet emission. We note that all these features hold even if a large, up to a noticeable fraction of the restoring beam, shift of one of the maps is applied in any direction. More typical $\alpha$ values are detected in the jet, at the core separation of about 10~mas and beyond, where influence of the scattered emission of the compact core feature becomes weaker. 

\subsection{Synchrotron self-absorbed spectrum of the core}
\label{s:synchrotron_spectrum}
\begin{figure}
    \centering
    \includegraphics[width=1\linewidth]{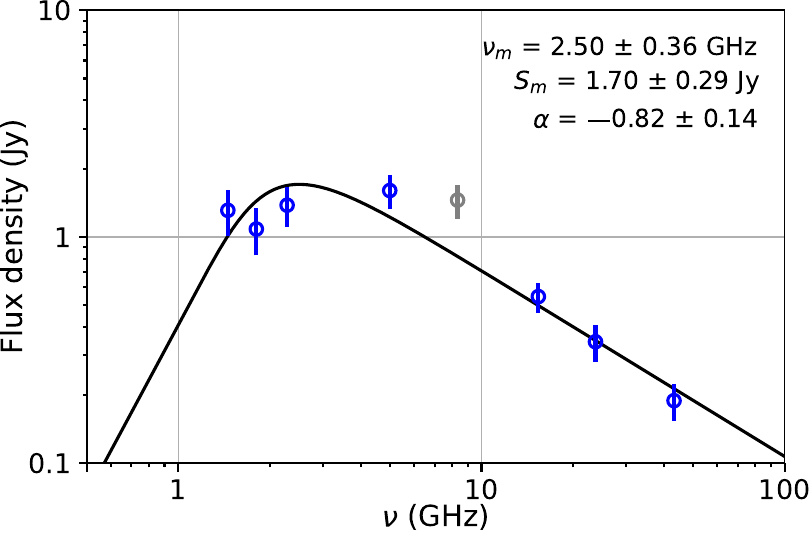}
    \caption{Spectral fit to the core component. Solid line represents the spectrum derived from the homogeneous synchrotron source model. Grey circle showing the 8.3~GHz data was not included in the fit as representing the most time separated epoch, at which the core flux density was likely affected by the internal source variability. The best-fit parameters are specified in the top right corner.}
    \label{fig:synch_spectrum}
\end{figure}

We fitted the spectrum of the measured core component flux density using the standard spectrum of a homogeneous incoherent synchrotron source of relativistic plasma \citep{Pacholczyk1970}

\begin{gather}
    S_{\nu} \propto \nu^{5/2} \left(1-\exp{\left[-\left(\frac{\nu_1}{\nu}\right)^{5/2 - \alpha}\right]}\right)\,, 
	%\label{eq:alpha_err},
\end{gather} 

\noindent where $\nu_1$ is the frequency at which the optical depth is $\tau=1$, $\alpha$ is the optically thin spectral index. The fitted spectrum is presented in \autoref{fig:synch_spectrum}. We excluded the measurement at 8.3~GHz from this analysis (grey circle in \autoref{fig:synch_spectrum}) as the only core flux density estimate at the most time separated epoch. At frequencies below 5.0~GHz, we took the median flux densities of the core component. Best fit parameters derived are $\nu_1 = 2.08 \pm 0.38$~GHz, $\alpha = - 0.82 \pm 0.14$. The peak flux density at the corresponding self-absorption turnover frequency $\nu_m = 2.50 \pm 0.36$~GHz  is $S_m = 1.70\pm0.29$~Jy. Using these fitted parameters of the synchrotron spectrum, we estimate the magnetic field $B$ within the source adopting the standard synchrotron theory and assuming that the emission region is uniform and spherical. The perpendicular to the line of sight component of the magnetic field can be estimated following \cite{Marscher83}

\begin{gather}
    B = 10^{-5}b(\alpha)\,\theta_m^4\,\nu_m^5\,S_m^{\prime-2} \left(\frac{\delta}{1+z}\right) [G]\,,
	\label{eq:magnetic_field}
\end{gather} 

\noindent where $\delta$ is the Doppler factor, $z$ is the redshift, $S_m^\prime = 2.21$~Jy is the flux density at $\nu_m$ extrapolated back from the straight-line optically thin slope, $\theta_m$ is the intrinsic angular size of the spherical component at the turnover frequency and $b(\alpha)$ is calculated following \cite{Pushkarev2019} as a function of spectral index $\alpha$, optical depth $\tau_m$ at $\nu_m$, physical constants and a conversion factor, which allows one to express $\nu_m$ in GHz, angular size $\theta_m$ in mas, and $S_m^\prime$ in Jy. We calculated $\theta_m$ as $1.8\theta_i \nu_m^{-1} = 3.02\pm1.09$~mas, where $\theta_i$ is the intrinsic size of the quasar at 1~GHz (see \autoref{s:k_scat_fit}) and the multiplicative factor of approximately 1.8 is used in order to make transition to spherical emission region shape \citep[e.g.][]{Marscher1987,Homan21}. Therefore, Eq.~\ref{eq:magnetic_field} yields the magnetic field for the apparent core at the turnover frequency $0.06\pm0.10$~G, which is comparable to estimates inferred for other parsec-scale AGN jets \citep[e.g.][]{Savolainen06,Pushkarev2019}.

\section{OVRO 15 GHz light curve analysis}
\label{s:lc}
Depending on the power-law index of spatial power spectrum of free-electron density fluctuations of plasma, both diffractive and refractive scattering effects can be observed \citep{Narayan1985,Cordes86,Armstrong95}. It is believed that unusually strong refractive effects in the ISM are responsible for Extreme Scattering Event \citep{Fiedler1987, Fiedler1994} and associated with a passage of localised structures of ionized medium with the AU-scale transverse dimensions across the line of sight \citep{Cordes86,Clegg1988}. ESEs show great variety in form and amplitude of the event on light curves, but a common feature is a pair of peaks (or caustics) separated by weeks or months, surrounding a flat or rounded minimum \citep{Fiedler1994,Pushkarev2013}. This variety is associated with the difference in the relative angular sizes between a background compact extragalactic source, scattering lens and angular broadening of a source.    

\cite{Fiedler1987} reported on the first detection of ESE in the light curves of the quasars 0954$+$658, 1502$+$106 and 1611$+$343 at 2.7~GHz. It has been shown that at 2.7~GHz flux density variations take a maximum deviation from the mean from 6~per~cent to 100~per~cent \citep{Fiedler1994}. \cite{Pushkarev2013} reported on the first serendipitous detection of the theoretically-predicted phenomenon of multiple parsec-scale imaging of the quasar 2023$+$334 induced by refractive effects when the source was undergoing an ESE.   

\begin{figure*}
    \centering
    \includegraphics[width=1\linewidth]{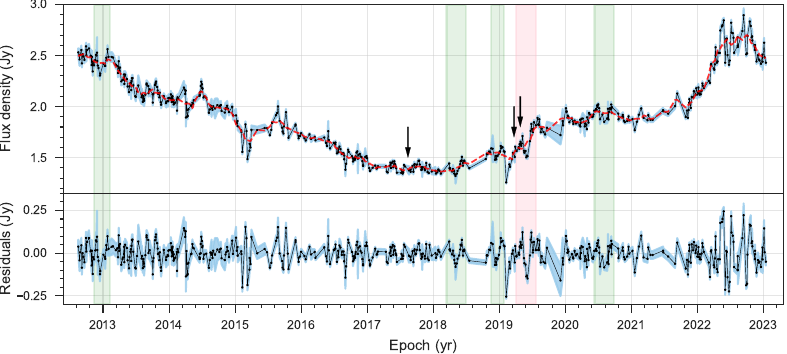}
    \caption{Upper panel: Full period 15~GHz OVRO light curve of the quasar 2005$+$403. The blue shaded area corresponds to a significance level of $1\sigma$. The red dotted curve was obtained by averaging the flux density over a range of $\pm0.1$~yr from each point of the original light curve and used for the long-term trend correction. Bottom panel: The OVRO light curve after the long-term trend correction. The pink rectangle indicates the time period when the source was undergoing the ESE modeled in \autoref{fig:2005+403_lc_mod}. The green rectangles show time ranges with other possible candidates to lower amplitude ESEs. The arrows point to the epochs at which secondary images of the source were detected (see \autoref{s:mapping} for details).}
    \label{fig:J2007_full_lc}
\end{figure*}
The formation of secondary images is believed to occur due to the crossing of rays refracted by the scattering screen at different angles for the observer \citep{Clegg1988}. These refracted rays may create caustic surfaces on the light curve of the background source. The flux density excursions during the scattering event provide valuable information for studying the physical properties of the intervening scattering screen. For the quasar 2005$+$403, we have found multiple episodes of formation of sub-images at different epochs and frequencies (see in \autoref{s:mapping}). To further investigate this phenomenon, the OVRO 15~GHz light curve of the quasar was analysed to search for characteristic flux density changes associated with extreme scattering events. The full OVRO light curve covers observational epochs from August 17, 2012 to January 14, 2023 (\autoref{fig:J2007_full_lc}, upper panel). It clearly shows the long-term flux density changes caused by the internal variability of the source. To simplify the search and analysis of ESE events, a boxcar averaging technique was applied with a time window of $\pm0.1$~yr (\autoref{fig:J2007_full_lc}, bottom panel). This width was chosen based on minimizing the average deviation of the measured data from the model curve. Finally, we subtracted the obtained average values from the original light curve data. 

The OVRO light curve of quasar 2005$+$403 is characterized by variations at different levels of amplitudes and time scales. After averaging and subtracting the long-term variations caused by intrinsic changes of the source, the light curve residuals better reveal short-term excursions in flux density (\autoref{fig:J2007_full_lc}, bottom panel). One of them, at epoch around 2019.42 clearly shows specific variations attributed to ESE (pink rectangle in \autoref{fig:J2007_full_lc}). The quasar appears to have undergone an ESE started around March 2019 (\autoref{fig:2005+403_lc_mod}). The duration of the event is approximately 1.6 months and the maximum flux density variation reaches 10~per~cent. The fact that this effect is detectable as an ESE even at the relatively high frequency of 15~GHz suggests that the refractive strength of the scattering lens during this event must have been high. It is worth noting that there are four more ESE candidates with somewhat smaller amplitude changes but comparable durations marked by the green stripes in \autoref{fig:J2007_full_lc}, indicating that the scattering screen moving across the line of sight towards the source is highly turbulent and inhomogeneous.

\subsection{Stochastic broadening model and properties of the screen}
\begin{figure}
    \centering
    \includegraphics[width=1\linewidth]{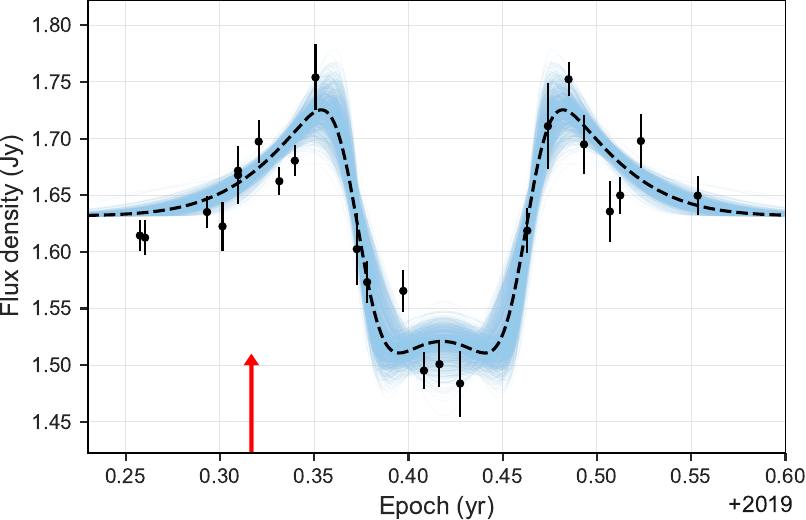}
    \caption{The 15~GHz OVRO light curve of the quasar 2005$+$403 after a long-term trend correction increased by the median value of the flux density for this period (1.6~Jy), assumed as an extreme scattering event (black dots). The dotted curve represents the best-fit line. Blue faint curves are the 1000 bootstrap fitted simulations used for error estimations. The red arrow sets the epoch (March 24, 2019), when two symmetrical sub-images of the core were detected at 1.4~GHz (project code BH222E).}
    \label{fig:2005+403_lc_mod}
\end{figure}

For a quantitative description of interstellar plasma lens properties we fitted the observed flux density variations on the light curve using the Fiedler's statistical model of flux redistribution \citep[see Appendix~A in][]{Fiedler1994}. The Fiedler's model suggests that the lens can be approximated by a single one-dimensional band-shaped lens projected on the sky with angular width $\theta_l$. It is assumed that the background source has a circular Gaussian brightness distribution with a FWHM $\theta_s$ and its observed angular width broadened into a Gaussian brightness distribution with FWHM $\theta_b$ \citep[see Figure~5 in][]{Fiedler1994}. Another fitted parameter is a proper motion $\mu$~-- the angular change in position of a lens across the line of sight, measured in mas per year. We applied a similar approach as in \cite{Pushkarev2013} using a two-component model, a lensed component with $\mathrm{S}_\mathrm{scat}$ and an unlensed component with $\mathrm{S}_\mathrm{unscat}$, contributing to the observed flux density. Their sum equals the nominal flux density level outside an ESE. Therefore, we fitted the measured flux density light curve in the period from March 2019 to June 2019 with five free parameters ($\theta_\mathrm{l}/\theta_\mathrm{s}$, $\theta_\mathrm{b}/\theta_\mathrm{s}$, $\mu/\theta_\mathrm{s}$, $\mathrm{S}_\mathrm{scat}$ and $\mathrm{S}_\mathrm{unscat}$) using non-linear least squares according to Eq. A5 from \cite{Fiedler1994}. The value of $\theta_\mathrm{b} = 0.48\pm0.05$~mas was determined as median angular size of the VLBI core using all the archival MOJAVE VLBA\footnote{\url{https://www.cv.nrao.edu/MOJAVE/sourcepages/2005+403.shtml}} data of 2005$+$403. The errors of the fitted parameters were determined by the bootstrap method. The results of fitting are illustrated in \autoref{fig:2005+403_lc_mod} and listed in \autoref{tab:lc_modelling0}. According to the model, the source was covered with a plasma lens for 1.61 months and the epoch of minimum flux density of the light curve during the ESE event is 2019.42 (middle value between peaks). The angular width of the scattering lens $\theta_\mathrm{l}=0.39\pm0.29$~mas. It exceeds the intrinsic angular size of the source $\theta_\mathrm{s}$, which equals $0.11\pm0.06$~mas. Scattering broadened it to the size of $\theta_\mathrm{b} = 0.48\pm0.05$~mas. The proper motion of plasma lens across the line of sight is $\mu = 4.37\pm3.13$~mas~yr$^{-1}$ which corresponds to the transverse velocity of the lens with respect to the observer $V_l = 4.74\,\mu D = 37.1\pm26.7$~km~s$^{-1}$, where $D$ is the distance to the lens in kpc. 

\cite{Rygl2012} measured the distances to the five massive star-forming regions towards the Cygnus~X complex within a 10\% accuracy. The angular separation between the average position of these regions and the quasar 2005$+$403 is $6\fdg1$ (where $\Delta l = 4\fdg8$ and $\Delta b = 3\fdg7$). For our calculations, we took the average value of the distances $D=1.8\pm0.1$~kpc. The correction of the transverse lens velocity due to the Earth's rotation around the Sun is approximately $-5.1$~km~s$^{-1}$ in the Sun's rest frame. Other alternative and more distant locations of the lens, such as the Perseus Arm or the Outer Arm are less likely. First, it would result in the transverse velocity values that are comparable with the speed of the Galactic rotation relative to the center at those separations. Second, a more remote screen is less effective. The transverse linear size of plasma lens $a = D\theta_\mathrm{l} = 0.7\pm0.5$~AU. We summarize all physical parameters of the scattering screen in \autoref{tab:lens_properties0}. Notably, the quasar 2005$+$403 was observed at 1.4~GHz as a phase calibrator (definitely, not the best choice as its short 30~sec scans prevented fringe detections due to heavy scattering resulted in a dramatic drop of the correlated flux density at middle and long baseline projections) for the VLBA experiment BH222E carried out on March 24, 2019, within the left caustic spike of the ESE with the total VLBI flux 2.24~Jy. The restored map at 1.4~GHz shows two symmetrical sub-images of the source formed nearly along the line of the constant Galactic latitude.

\begin{table}
	\caption{The OVRO light curve modelling results.}
	\centering
	\begin{tabular}{|*{3}{c|}} 
		\hline\hline
		   Parameter   & Result  & Unit\\
		\hline
		Min. epoch$^{*}$                            & 2019.42        & yr  \\
		$\mathrm{\tau}$$^{*}$                       & 0.13           & yr  \\
		$\mathrm{\theta_{b}}$$^{*}$                 & $0.48\pm 0.05$ & mas \\
		$\mathrm{\mu}$/$\mathrm{\theta_{s}}$        & $40.1\pm 16.9$ & yr$^{-1}$\\
		$\mathrm{\theta_{l}}$/$\mathrm{\theta_{s}}$ & $3.58\pm 1.61$ & \\
		$\mathrm{\theta_{b}}$/$\mathrm{\theta_{s}}$ & $4.41\pm 2.51$ & \\
		$\mathrm{S}_\mathrm{scat}$                  & $0.31\pm 0.11$ & Jy \\
		$\mathrm{S}_\mathrm{unscat}$                & $1.32\pm 0.11$ & Jy \\
		\hline
	\end{tabular}
		\begin{tablenotes}
            \item $^{*}$These parameters are not fitted.
        \end{tablenotes}
\label{tab:lc_modelling0}
\end{table}

\begin{table}
	\caption{The summary of plasma lens properties.}
	\centering
	\begin{tabular}{|*{3}{c|}} % four columns, alignment for each
		\hline\hline
		   Parameter   & Result &  Unit \\
		\hline
		$\mathrm{\mu}$           &  $ 4.37\pm3.13$  & mas yr $^{-1}$\\
		%$\mathrm{\theta_{s}}$    &  $ 0.11\pm0.06 $ & mas\\
		$\mathrm{\theta_{l}}$    &  $ 0.39\pm0.29$  & mas\\
		$\mathrm{a}$             &  $ 0.70\pm 0.51$  & AU \\
		$\mathrm{V_l}$           &  $ 37.1\pm 26.7$ & km s$^{-1}$\\
		\hline
	\end{tabular}
	\label{tab:lens_properties0}
\end{table}

The quasar 2005$+$403 is seen through a particularly dense and inhomogeneous part of the ISM of the Galaxy. The main feature of this source is the occurrence of multiple imaging events often seen on the low-frequency VLBI maps. However, not all of these multiple imaging events at frequencies $\lesssim$5~GHz can be directly related to characteristic caustic surfaces on the 15.0~GHz OVRO light curve (see \autoref{fig:J2007_full_lc} in July 12, 2017). One likely reason is the scattering strength of the intervening lens, the corresponding splitting angle of which decreases faster ($\propto\nu^{-2}$) with observing frequency compared to the angular resolution ($\propto\nu^{-1}$). As a result, the effect that is consistently observed in low-frequency (1.4--5.0~GHz) VLBA maps may be too weak for robust detection at a 15~GHz light curve. Another possibility is that the multi-component structure of the background source, as well as the clumpy inhomogeneous structure of the scattering screen are able to induce more than one ESE in a row on the light curve, making the identification of individual ESEs more challenging, as they may partially overlap each other. It has been discussed that a double lens system can explain the observed multiple ESEs (two or tree in a row) on a light curve \citep{Hee2005}.

\section{Summary}
\label{s:summary}
The quasar 2005$+$403, located at a low Galactic latitude ($b=4\fdg3$), manifests a typical one-sided core-jet structure on parsec scales probed by VLBA observations at frequencies 43.2, 23.8, 15.4, 8.3 and 5.0~GHz. However, at lower frequencies (8.3, 5.0, 2.3, 1.8 and 1.4~GHz) it shows strong evidence of anisotropic scattering, making it the second AGN \citep[after the quasar 2023$+$335 by][]{Pushkarev2013} with established refractive-dominated scattering based on the results of the joint analysis of the interferometric VLBA and single-dish OVRO observations.

The scattering-induced patterns, formed by multiple imaging of the compact core, extend along the position angle of about $40^\circ$ at different frequencies (8.3~GHz and below) and observing epochs, corresponding to a line of constant Galactic latitude. Analyzing non-simultaneous multi-frequency VLBA observations of the source, we found that the angular separation between the core and its secondary images follows a wavelength squared dependence, supporting that the patterns are caused by scattering in the interstellar medium. The angular size of the VLBI core scales as $\nu^{-k}$, where $k=2.00\pm0.08$, if the data at all available frequencies ranging from 0.6 to 43.2~GHz are fitted. The value of $k$ agrees within the errors with the canonical 2.2 expected for Kolmogorov turbulence, when only the low-frequency ($\lesssim8.3$~GHz) data are considered. Extrapolating this dependence to the frequencies of LOFAR and SKA, the expected angular size of the quasar to be observed is at the level of tens of arcseconds, roughly 100 times larger than its intrinsic unscattered extent. By fitting a synchrotron spectrum to the core component flux densities, the magnetic field in the core was assessed to be at a level of 0.06~G. The progressively stronger scattering at low frequencies leads to spectral index distributions with extreme $\alpha$-values, e.g. ranging from $-6$ to $+2$ for the simultaneous VLBA observations at 2.3 and 5.0~GHz.

The 15~GHz OVRO light curve of the quasar 2005$+$403 revealed a distinct extreme scattering event (ESE) with flux density variations at the level of about 10 per cent. This event occurred around the epoch 2019.42 and lasted for nearly 1.6 months. By fitting the ESE with a flux redistribution model based on stochastic broadening \citep{Fiedler1987}, we inferred that the scattering screen responsible for the event has a transverse angular size of about 0.4~mas and drifts across the line of sight with a proper motion of 4.4~mas~yr$^{-1}$. Assuming that the lens is located in the Cygnus region at a distance of 1.8~kpc, its transverse linear size and transverse speed with respect to the observer are 0.7~AU and 31~km~s$^{-1}$, respectively. The direction of the proper motion is prograde with regard to the Galactic center, which corresponds to a direction from North-East to South-West on the maps in the equatorial coordinates. Additionally, the light curve showed several more potential ESEs with smaller flux density amplitude variability but comparable duration. Three of these events are consecutive, suggesting a complex structure of the scattering screen. The simultaneous detection of the sub-images in opposite directions from the primary core image, typically at low frequencies (from 1.4 to 2.3~GHz), further indicates that the line of sight toward the source often crosses at least two screens, each creating the edge effect of multiple imaging. This is consistent with the long-term light curve of 2005$+$403 taken with the RATAN-600 at 2~GHz, which does not show characteristic variability attributed to a single ESE (S.Trushkin, priv. comm.). The fact that the refraction-induced patterns are stretched out along the line of constant Galactic latitude reflects not only the orbital motion of the lens system but also suggests that the discrete structures of free-electron density enhancements in the scattering screen have a close-to-flat geometry, which might represent shock fronts. Understanding the exact nature of these scattering plasma lenses in the intervening medium requires further investigation and more detailed modeling. 

The unique properties of the quasar 2005$+$403 in terms of scattering effects and its large angular size in the low-frequency domain make it one of the most promising sources for further studying propagation phenomena. In particular, observations with LOFAR and SKA will provide valuable insights into these processes and contribute to our understanding of turbulence in the Galactic ISM.

\section*{Acknowledgements}
We thank the anonymous referee for useful comments, which helped to improve the manuscript.
This study was supported by the Russian Science Foundation grant\footnote{Information about the project: \url{https://rscf.ru/en/project/21-12-00241/}} 21-12-00241.
This research has made use of data from the OVRO 40-m monitoring program \citep{Richards2011}, supported by private funding from the California Institute of Technology and the Max Planck Institute for Radio Astronomy, and by NASA grants NNX08AW31G, NNX11A043G, and NNX14AQ89G and NSF grants AST-0808050, AST1109911, and AST-1835400.
This research has used the MOJAVE database maintained by the MOJAVE team \citep{2018ApJS..234...12L}. 
This work made use of the Swinburne University of Technology software correlator, developed as part of the Australian Major National Research Facilities Programme and operated under licence.
S.K. acknowledges support from the European Research Council (ERC) under the European Unions Horizon 2020 research and innovation programme under grant agreement No.~771282.

%%%%%%%%%%%%%%%%%%%%%%%%%%%%%%%%%%%%%%%%%%%%%%%%%%
\section*{Data Availability}
In this study, we used the data from the VLBA experiments with project codes BF025A, BG196H, BG246T, BG258G, BH222E, BP240A, BP240C and MOJAVE sessions. See \autoref{s:data} for details. The original data is available at the NRAO \href{https://data.nrao.edu/}{website}.
The OVRO 40m light curve underlying this article will be shared on reasonable request to the OVRO 40m Monitoring collaboration.

%%%%%%%%%%%%%%%%%%%% REFERENCES %%%%%%%%%%%%%%%%%%

\bibliographystyle{mnras}
\bibliography{article}

\begin{thebibliography}{}
\makeatletter
\relax
\def\mn@urlcharsother{\let\do\@makeother \do\$\do\&\do\#\do\^\do\_\do\%\do\~}
\def\mn@doi{\begingroup\mn@urlcharsother \@ifnextchar [ {\mn@doi@}
  {\mn@doi@[]}}
\def\mn@doi@[#1]#2{\def\@tempa{#1}\ifx\@tempa\@empty \href
  {http://dx.doi.org/#2} {doi:#2}\else \href {http://dx.doi.org/#2} {#1}\fi
  \endgroup}
\def\mn@eprint#1#2{\mn@eprint@#1:#2::\@nil}
\def\mn@eprint@arXiv#1{\href {http://arxiv.org/abs/#1} {{\tt arXiv:#1}}}
\def\mn@eprint@dblp#1{\href {http://dblp.uni-trier.de/rec/bibtex/#1.xml}
  {dblp:#1}}
\def\mn@eprint@#1:#2:#3:#4\@nil{\def\@tempa {#1}\def\@tempb {#2}\def\@tempc
  {#3}\ifx \@tempc \@empty \let \@tempc \@tempb \let \@tempb \@tempa \fi \ifx
  \@tempb \@empty \def\@tempb {arXiv}\fi \@ifundefined
  {mn@eprint@\@tempb}{\@tempb:\@tempc}{\expandafter \expandafter \csname
  mn@eprint@\@tempb\endcsname \expandafter{\@tempc}}}

\bibitem[\protect\citeauthoryear{{Armstrong}, {Rickett}  \&
  {Spangler}}{{Armstrong} et~al.}{1995}]{Armstrong95}
{Armstrong} J.~W.,  {Rickett} B.~J.,   {Spangler} S.~R.,  1995, \mn@doi [\apj]
  {10.1086/175515}, \href
  {https://ui.adsabs.harvard.edu/abs/1995ApJ...443..209A} {443, 209}

\bibitem[\protect\citeauthoryear{{Baars}, {Genzel}, {Pauliny-Toth}  \&
  {Witzel}}{{Baars} et~al.}{1977}]{1977A&A....61...99B}
{Baars} J.~W.~M.,  {Genzel} R.,  {Pauliny-Toth} I.~I.~K.,   {Witzel} A.,  1977,
  \aap, \href {https://ui.adsabs.harvard.edu/abs/1977A&A....61...99B} {61, 99}

\bibitem[\protect\citeauthoryear{{Beuther}, {Schilke}, {Menten}, {Motte},
  {Sridharan}  \& {Wyrowski}}{{Beuther} et~al.}{2002}]{Beuther2002}
{Beuther} H.,  {Schilke} P.,  {Menten} K.~M.,  {Motte} F.,  {Sridharan} T.~K.,
   {Wyrowski} F.,  2002, \mn@doi [\apj] {10.1086/338334}, \href
  {https://ui.adsabs.harvard.edu/abs/2002ApJ...566..945B} {566, 945}

\bibitem[\protect\citeauthoryear{{Blandford} \& {K{\"o}nigl}}{{Blandford} \&
  {K{\"o}nigl}}{1979}]{BK79}
{Blandford} R.~D.,  {K{\"o}nigl} A.,  1979, \mn@doi [\apj] {10.1086/157262},
  \href {https://ui.adsabs.harvard.edu/abs/1979ApJ...232...34B} {232, 34}

\bibitem[\protect\citeauthoryear{{Boksenberg}, {Briggs}, {Carswell}, {Schmidt}
  \& {Walsh}}{{Boksenberg} et~al.}{1976}]{Boksenberg76}
{Boksenberg} A.,  {Briggs} S.~A.,  {Carswell} R.~F.,  {Schmidt} M.,   {Walsh}
  D.,  1976, \mn@doi [\mnras] {10.1093/mnras/177.1.43P}, \href
  {https://ui.adsabs.harvard.edu/abs/1976MNRAS.177P..43B} {177, 43}

\bibitem[\protect\citeauthoryear{{Clegg}, Chernoff  \& Cordes}{{Clegg}
  et~al.}{1988}]{Clegg1988}
{Clegg} A.~W.,  Chernoff D.,   Cordes J.,  1988, in {Cordes} J.~M.,  Rickett
  B.~J.,   Backer D.~C.,  eds,  American Institute of Physics Conference Series
  Vol. 174, Radio Wave Scattering in the Interstellar Medium. pp 174--178,
  \mn@doi{10.1063/1.37587}

\bibitem[\protect\citeauthoryear{{Clegg}, {Fey}  \& {Lazio}}{{Clegg}
  et~al.}{1998}]{Clegg_1998}
{Clegg} A.~W.,  {Fey} A.~L.,   {Lazio} T. J.~W.,  1998, \mn@doi [\apj]
  {10.1086/305344}, \href
  {https://ui.adsabs.harvard.edu/abs/1998ApJ...496..253C} {496, 253}

\bibitem[\protect\citeauthoryear{{Cordes}, {Pidwerbetsky}  \&
  {Lovelace}}{{Cordes} et~al.}{1986}]{Cordes86}
{Cordes} J.~M.,  {Pidwerbetsky} A.,   {Lovelace} R.~V.~E.,  1986, \mn@doi
  [\apj] {10.1086/164728}, \href
  {https://ui.adsabs.harvard.edu/abs/1986ApJ...310..737C} {310, 737}

\bibitem[\protect\citeauthoryear{{Cordes}, Rickett  \& Backer}{{Cordes}
  et~al.}{1988}]{Cordes1988}
{Cordes} J.~M.,  Rickett B.~J.,   Backer D.~C.,  1988, in Radio Wave Scattering
  in the Interstellar Medium.

\bibitem[\protect\citeauthoryear{{Cyganowski}, {Reid}, {Fish}  \&
  {Ho}}{{Cyganowski} et~al.}{2003}]{Cyganowski2003}
{Cyganowski} C.~J.,  {Reid} M.~J.,  {Fish} V.~L.,   {Ho} P.~T.~P.,  2003,
  \mn@doi [\apj] {10.1086/377688}, \href
  {https://ui.adsabs.harvard.edu/abs/2003ApJ...596..344C} {596, 344}

\bibitem[\protect\citeauthoryear{{Deller} et~al.,}{{Deller}
  et~al.}{2011}]{DifX}
{Deller} A.~T.,  et~al., 2011, \mn@doi [\pasp] {10.1086/658907}, \href
  {https://ui.adsabs.harvard.edu/abs/2011PASP..123..275D} {123, 275}

\bibitem[\protect\citeauthoryear{{Desai} \& {Fey}}{{Desai} \&
  {Fey}}{2001}]{Desai2001}
{Desai} K.~M.,  {Fey} A.~L.,  2001, \mn@doi [\apjs] {10.1086/320349}, \href
  {https://ui.adsabs.harvard.edu/abs/2001ApJS..133..395D} {133, 395}

\bibitem[\protect\citeauthoryear{{Downes} \& {Rinehart}}{{Downes} \&
  {Rinehart}}{1966}]{Downes1966}
{Downes} D.,  {Rinehart} R.,  1966, \mn@doi [\apj] {10.1086/148691}, \href
  {https://ui.adsabs.harvard.edu/abs/1966ApJ...144..937D} {144, 937}

\bibitem[\protect\citeauthoryear{{Fey}, {Spangler}  \& {Mutel}}{{Fey}
  et~al.}{1989}]{Fey1989}
{Fey} A.~L.,  {Spangler} S.~R.,   {Mutel} R.~L.,  1989, \mn@doi [\apj]
  {10.1086/167144}, \href
  {https://ui.adsabs.harvard.edu/abs/1989ApJ...337..730F} {337, 730}

\bibitem[\protect\citeauthoryear{{Fiedler}, {Dennison}, {Johnston}  \&
  {Hewish}}{{Fiedler} et~al.}{1987}]{Fiedler1987}
{Fiedler} R.~L.,  {Dennison} B.,  {Johnston} K.~J.,   {Hewish} A.,  1987,
  \mn@doi [\nat] {10.1038/326675a0}, \href
  {https://ui.adsabs.harvard.edu/abs/1987Natur.326..675F} {326, 675}

\bibitem[\protect\citeauthoryear{{Fiedler}, {Dennison}, {Johnston}, {Waltman}
  \& {Simon}}{{Fiedler} et~al.}{1994}]{Fiedler1994}
{Fiedler} R.,  {Dennison} B.,  {Johnston} K.~J.,  {Waltman} E.~B.,   {Simon}
  R.~S.,  1994, \mn@doi [\apj] {10.1086/174432}, \href
  {https://ui.adsabs.harvard.edu/abs/1994ApJ...430..581F} {430, 581}

\bibitem[\protect\citeauthoryear{{Gab{\'a}nyi}, {Krichbaum}, {Britzen}, {Bach},
  {Ros}, {Witzel}  \& {Zensus}}{{Gab{\'a}nyi} et~al.}{2006}]{Gabani2006}
{Gab{\'a}nyi} K.~{\'E}.,  {Krichbaum} T.~P.,  {Britzen} S.,  {Bach} U.,  {Ros}
  E.,  {Witzel} A.,   {Zensus} J.~A.,  2006, \mn@doi [\aap]
  {10.1051/0004-6361:20054017}, \href
  {https://ui.adsabs.harvard.edu/abs/2006A&A...451...85G} {451, 85}

\bibitem[\protect\citeauthoryear{{Goodman} \& {Narayan}}{{Goodman} \&
  {Narayan}}{1985}]{Narayan1985}
{Goodman} J.,  {Narayan} R.,  1985, \mn@doi [\mnras] {10.1093/mnras/214.4.519},
  \href {https://ui.adsabs.harvard.edu/abs/1985MNRAS.214..519G} {214, 519}

\bibitem[\protect\citeauthoryear{{Greisen}}{{Greisen}}{2003}]{Greisen2003}
{Greisen} E.~W.,  2003, in {Heck} A.,  ed.,  Astrophysics and Space Science
  Library Vol. 285, Information Handling in Astronomy - Historical Vistas.
  p.~109, \mn@doi{10.1007/0-306-48080-8_7}

\bibitem[\protect\citeauthoryear{{H{\"o}gbom}}{{H{\"o}gbom}}{1974}]{Hongbom1974}
{H{\"o}gbom} J.~A.,  1974, \aaps, \href
  {https://ui.adsabs.harvard.edu/abs/1974A&AS...15..417H} {15, 417}

\bibitem[\protect\citeauthoryear{{Homan} et~al.,}{{Homan}
  et~al.}{2021}]{Homan21}
{Homan} D.~C.,  et~al., 2021, \mn@doi [\apj] {10.3847/1538-4357/ac27af}, \href
  {https://ui.adsabs.harvard.edu/abs/2021ApJ...923...67H} {923, 67}

\bibitem[\protect\citeauthoryear{{Hovatta} et~al.,}{{Hovatta}
  et~al.}{2014}]{HovattaAller2014}
{Hovatta} T.,  et~al., 2014, \mn@doi [\aj] {10.1088/0004-6256/147/6/143}, \href
  {https://ui.adsabs.harvard.edu/abs/2014AJ....147..143H} {147, 143}

\bibitem[\protect\citeauthoryear{{Jennison}}{{Jennison}}{1958}]{Jennison1958}
{Jennison} R.~C.,  1958, \mn@doi [\mnras] {10.1093/mnras/118.3.276}, \href
  {https://ui.adsabs.harvard.edu/abs/1958MNRAS.118..276J} {118, 276}

\bibitem[\protect\citeauthoryear{{Kellermann} \& {Pauliny-Toth}}{{Kellermann}
  \& {Pauliny-Toth}}{1981}]{KP81}
{Kellermann} K.~I.,  {Pauliny-Toth} I.~I.~K.,  1981, \mn@doi [\araa]
  {10.1146/annurev.aa.19.090181.002105}, \href
  {https://ui.adsabs.harvard.edu/abs/1981ARA&A..19..373K} {19, 373}

\bibitem[\protect\citeauthoryear{{Kellermann} et~al.,}{{Kellermann}
  et~al.}{2007}]{Kellermann07}
{Kellermann} K.~I.,  et~al., 2007, \mn@doi [\apss] {10.1007/s10509-007-9622-5},
  \href {https://ui.adsabs.harvard.edu/abs/2007Ap&SS.311..231K} {311, 231}

\bibitem[\protect\citeauthoryear{{Kim} \& {Lee}}{{Kim} \&
  {Lee}}{2005}]{Hee2005}
{Kim} H.~I.,  {Lee} H.~M.,  2005, J. Korean Phys. Soc., 47, 1064

\bibitem[\protect\citeauthoryear{{Kolmogorov}}{{Kolmogorov}}{1941}]{Kolmogorov41}
{Kolmogorov} A.,  1941, Proceedings of the USSR Academy of Sciences, pp
  299--303

\bibitem[\protect\citeauthoryear{{Konigl}}{{Konigl}}{1981}]{Koenigl81}
{Konigl} A.,  1981, \mn@doi [\apj] {10.1086/158638}, \href
  {https://ui.adsabs.harvard.edu/abs/1981ApJ...243..700K} {243, 700}

\bibitem[\protect\citeauthoryear{{Koryukova}, {Pushkarev}, {Plavin}  \&
  {Kovalev}}{{Koryukova} et~al.}{2022}]{Koryukova2022}
{Koryukova} T.~A.,  {Pushkarev} A.~B.,  {Plavin} A.~V.,   {Kovalev} Y.~Y.,
  2022, \mn@doi [\mnras] {10.1093/mnras/stac1898}, \href
  {https://ui.adsabs.harvard.edu/abs/2022MNRAS.515.1736K} {515, 1736}

\bibitem[\protect\citeauthoryear{{Kovalev} et~al.,}{{Kovalev}
  et~al.}{2005}]{Kovalev05}
{Kovalev} Y.~Y.,  et~al., 2005, \mn@doi [\aj] {10.1086/497430}, \href
  {https://ui.adsabs.harvard.edu/abs/2005AJ....130.2473K} {130, 2473}

\bibitem[\protect\citeauthoryear{{Lister}, {Aller}, {Aller}, {Hovatta},
  {Max-Moerbeck}, {Readhead}, {Richards}  \& {Ros}}{{Lister}
  et~al.}{2015}]{Lister15}
{Lister} M.~L.,  {Aller} M.~F.,  {Aller} H.~D.,  {Hovatta} T.,  {Max-Moerbeck}
  W.,  {Readhead} A.~C.~S.,  {Richards} J.~L.,   {Ros} E.,  2015, \mn@doi
  [\apjl] {10.1088/2041-8205/810/1/L9}, \href
  {https://ui.adsabs.harvard.edu/abs/2015ApJ...810L...9L} {810, L9}

\bibitem[\protect\citeauthoryear{{Lister}, {Aller}, {Aller}, {Hodge}, {Homan},
  {Kovalev}, {Pushkarev}  \& {Savolainen}}{{Lister}
  et~al.}{2018}]{2018ApJS..234...12L}
{Lister} M.~L.,  {Aller} M.~F.,  {Aller} H.~D.,  {Hodge} M.~A.,  {Homan} D.~C.,
   {Kovalev} Y.~Y.,  {Pushkarev} A.~B.,   {Savolainen} T.,  2018, \mn@doi
  [\apjs] {10.3847/1538-4365/aa9c44}, \href
  {https://ui.adsabs.harvard.edu/abs/2018ApJS..234...12L} {234, 12}

\bibitem[\protect\citeauthoryear{{Lister} et~al.,}{{Lister}
  et~al.}{2019}]{Lister19}
{Lister} M.~L.,  et~al., 2019, \mn@doi [\apj] {10.3847/1538-4357/ab08ee}, \href
  {https://ui.adsabs.harvard.edu/abs/2019ApJ...874...43L} {874, 43}

\bibitem[\protect\citeauthoryear{{Lobanov}}{{Lobanov}}{1998}]{Lobanov98}
{Lobanov} A.~P.,  1998, \aap, \href
  {https://ui.adsabs.harvard.edu/abs/1998A&A...330...79L} {330, 79}

\bibitem[\protect\citeauthoryear{{Lovelace}, {Salpeter}, {Sharp}  \&
  {Harris}}{{Lovelace} et~al.}{1970}]{Lovelace1970}
{Lovelace} R.~V.~E.,  {Salpeter} E.~E.,  {Sharp} L.~E.,   {Harris} D.~E.,
  1970, \mn@doi [\apj] {10.1086/150382}, \href
  {https://ui.adsabs.harvard.edu/abs/1970ApJ...159.1047L} {159, 1047}

\bibitem[\protect\citeauthoryear{{Marscher}}{{Marscher}}{1983}]{Marscher83}
{Marscher} A.~P.,  1983, \mn@doi [\apj] {10.1086/160597}, \href
  {https://ui.adsabs.harvard.edu/abs/1983ApJ...264..296M} {264, 296}

\bibitem[\protect\citeauthoryear{{Marscher}}{{Marscher}}{1987}]{Marscher1987}
{Marscher} A.~P.,  1987, in {Zensus} J.~A.,  {Pearson} T.~J.,  eds,
  Superluminal Radio Sources. pp 280--300

\bibitem[\protect\citeauthoryear{{Motte}, {Bontemps}, {Schilke}, {Schneider},
  {Menten}  \& {Brogui{\`e}re}}{{Motte} et~al.}{2007}]{Motte2007}
{Motte} F.,  {Bontemps} S.,  {Schilke} P.,  {Schneider} N.,  {Menten} K.~M.,
  {Brogui{\`e}re} D.,  2007, \mn@doi [\aap] {10.1051/0004-6361:20077843}, \href
  {https://ui.adsabs.harvard.edu/abs/2007A&A...476.1243M} {476, 1243}

\bibitem[\protect\citeauthoryear{{Pacholczyk}}{{Pacholczyk}}{1970}]{Pacholczyk1970}
{Pacholczyk} A.~G.,  1970, {Radio astrophysics. Nonthermal processes in
  galactic and extragalactic sources}

\bibitem[\protect\citeauthoryear{{Planck Collaboration} et~al.,}{{Planck
  Collaboration} et~al.}{2020}]{cosmology2020}
{Planck Collaboration} et~al., 2020, \mn@doi [\aap]
  {10.1051/0004-6361/201833910}, \href
  {https://ui.adsabs.harvard.edu/abs/2020A&A...641A...6P} {641, A6}

\bibitem[\protect\citeauthoryear{{Pushkarev} \& {Kovalev}}{{Pushkarev} \&
  {Kovalev}}{2012}]{PushkarevKovalev2012}
{Pushkarev} A.~B.,  {Kovalev} Y.~Y.,  2012, \mn@doi [\aap]
  {10.1051/0004-6361/201219352}, \href
  {https://ui.adsabs.harvard.edu/abs/2012A&A...544A..34P} {544, A34}

\bibitem[\protect\citeauthoryear{{Pushkarev} et~al.,}{{Pushkarev}
  et~al.}{2013}]{Pushkarev2013}
{Pushkarev} A.~B.,  et~al., 2013, \mn@doi [\aap] {10.1051/0004-6361/201321484},
  \href {https://ui.adsabs.harvard.edu/abs/2013A&A...555A..80P} {555, A80}

\bibitem[\protect\citeauthoryear{{Pushkarev}, {Butuzova}, {Kovalev}  \&
  {Hovatta}}{{Pushkarev} et~al.}{2019}]{Pushkarev2019}
{Pushkarev} A.~B.,  {Butuzova} M.~S.,  {Kovalev} Y.~Y.,   {Hovatta} T.,  2019,
  \mn@doi [\mnras] {10.1093/mnras/sty2724}, \href
  {https://ui.adsabs.harvard.edu/abs/2019MNRAS.482.2336P} {482, 2336}

\bibitem[\protect\citeauthoryear{{Readhead}, {Lawrence}, {Myers}, {Sargent},
  {Hardebeck}  \& {Moffet}}{{Readhead} et~al.}{1989}]{1989ApJ...346..566R}
{Readhead} A.~C.~S.,  {Lawrence} C.~R.,  {Myers} S.~T.,  {Sargent} W.~L.~W.,
  {Hardebeck} H.~E.,   {Moffet} A.~T.,  1989, \mn@doi [\apj] {10.1086/168039},
  \href {https://ui.adsabs.harvard.edu/abs/1989ApJ...346..566R} {346, 566}

\bibitem[\protect\citeauthoryear{{Readhead} et~al.,}{{Readhead}
  et~al.}{2021}]{Readhead21}
{Readhead} A.~C.~S.,  et~al., 2021, \mn@doi [\apj] {10.3847/1538-4357/abd08c},
  \href {https://ui.adsabs.harvard.edu/abs/2021ApJ...907...61R} {907, 61}

\bibitem[\protect\citeauthoryear{{Richards} et~al.,}{{Richards}
  et~al.}{2011}]{Richards2011}
{Richards} J.~L.,  et~al., 2011, \mn@doi [\apjs] {10.1088/0067-0049/194/2/29},
  \href {https://ui.adsabs.harvard.edu/abs/2011ApJS..194...29R} {194, 29}

\bibitem[\protect\citeauthoryear{{Rickett}}{{Rickett}}{1977}]{Rickett77}
{Rickett} B.~J.,  1977, \mn@doi [\araa] {10.1146/annurev.aa.15.090177.002403},
  \href {https://ui.adsabs.harvard.edu/abs/1977ARA&A..15..479R} {15, 479}

\bibitem[\protect\citeauthoryear{{Romani}, {Blandford}  \& {Cordes}}{{Romani}
  et~al.}{1987}]{Romani1987}
{Romani} R.~W.,  {Blandford} R.~D.,   {Cordes} J.~M.,  1987, \mn@doi [\nat]
  {10.1038/328324a0}, \href
  {https://ui.adsabs.harvard.edu/abs/1987Natur.328..324R} {328, 324}

\bibitem[\protect\citeauthoryear{{Rygl} et~al.,}{{Rygl}
  et~al.}{2012}]{Rygl2012}
{Rygl} K.~L.~J.,  et~al., 2012, \mn@doi [\aap] {10.1051/0004-6361/201118211},
  \href {https://ui.adsabs.harvard.edu/abs/2012A&A...539A..79R} {539, A79}

\bibitem[\protect\citeauthoryear{{Savolainen}, {Wiik}, {Valtaoja}  \&
  {Tornikoski}}{{Savolainen} et~al.}{2006}]{Savolainen06}
{Savolainen} T.,  {Wiik} K.,  {Valtaoja} E.,   {Tornikoski} M.,  2006, \mn@doi
  [\aap] {10.1051/0004-6361:20053753}, \href
  {https://ui.adsabs.harvard.edu/abs/2006A&A...446...71S} {446, 71}

\bibitem[\protect\citeauthoryear{{Schinzel}, {Lobanov}, {Taylor}, {Jorstad},
  {Marscher}  \& {Zensus}}{{Schinzel} et~al.}{2012}]{Schinzel2012}
{Schinzel} F.~K.,  {Lobanov} A.~P.,  {Taylor} G.~B.,  {Jorstad} S.~G.,
  {Marscher} A.~P.,   {Zensus} J.~A.,  2012, \mn@doi [\aap]
  {10.1051/0004-6361/201117705}, \href
  {https://ui.adsabs.harvard.edu/abs/2012A&A...537A..70S} {537, A70}

\bibitem[\protect\citeauthoryear{{Shepherd}}{{Shepherd}}{1997}]{Shepherd_1997}
{Shepherd} M.~C.,  1997, in {Hunt} G.,  {Payne} H.,  eds,  Astronomical Society
  of the Pacific Conference Series Vol. 125, Astronomical Data Analysis
  Software and Systems VI. p.~77

\bibitem[\protect\citeauthoryear{{Sridharan}, {Beuther}, {Schilke}, {Menten}
  \& {Wyrowski}}{{Sridharan} et~al.}{2002}]{Sridharan2002}
{Sridharan} T.~K.,  {Beuther} H.,  {Schilke} P.,  {Menten} K.~M.,   {Wyrowski}
  F.,  2002, \mn@doi [\apj] {10.1086/338332}, \href
  {https://ui.adsabs.harvard.edu/abs/2002ApJ...566..931S} {566, 931}

\bibitem[\protect\citeauthoryear{{Twiss}, {Carter}  \& {Little}}{{Twiss}
  et~al.}{1960}]{Twiss1960}
{Twiss} R.~Q.,  {Carter} A.~W.~L.,   {Little} A.~G.,  1960, The Observatory,
  \href {https://ui.adsabs.harvard.edu/abs/1960Obs....80..153T} {80, 153}

\bibitem[\protect\citeauthoryear{{Uyan{\i}ker}, {F{\"u}rst}, {Reich},
  {Aschenbach}  \& {Wielebinski}}{{Uyan{\i}ker} et~al.}{2001}]{Uyaniker2001}
{Uyan{\i}ker} B.,  {F{\"u}rst} E.,  {Reich} W.,  {Aschenbach} B.,
  {Wielebinski} R.,  2001, \mn@doi [\aap] {10.1051/0004-6361:20010387}, \href
  {https://ui.adsabs.harvard.edu/abs/2001A&A...371..675U} {371, 675}

\bibitem[\protect\citeauthoryear{{Vedantham} et~al.,}{{Vedantham}
  et~al.}{2017a}]{Vedantham2017b}
{Vedantham} H.~K.,  et~al., 2017a, \mn@doi [\apj] {10.3847/1538-4357/aa745c},
  \href {https://ui.adsabs.harvard.edu/abs/2017ApJ...845...89V} {845, 89}

\bibitem[\protect\citeauthoryear{{Vedantham} et~al.,}{{Vedantham}
  et~al.}{2017b}]{Vedantham2017}
{Vedantham} H.~K.,  et~al., 2017b, \mn@doi [\apj] {10.3847/1538-4357/aa7741},
  \href {https://ui.adsabs.harvard.edu/abs/2017ApJ...845...90V} {845, 90}

\bibitem[\protect\citeauthoryear{{Wendker}, {Higgs}  \& {Landecker}}{{Wendker}
  et~al.}{1991}]{Wendker1991}
{Wendker} H.~J.,  {Higgs} L.~A.,   {Landecker} T.~L.,  1991, \aap, \href
  {https://ui.adsabs.harvard.edu/abs/1991A&A...241..551W} {241, 551}

\bibitem[\protect\citeauthoryear{{Wright}, {Drake}, {Drew}  \& {Vink}}{{Wright}
  et~al.}{2010}]{Wright2010}
{Wright} N.~J.,  {Drake} J.~J.,  {Drew} J.~E.,   {Vink} J.~S.,  2010, \mn@doi
  [\apj] {10.1088/0004-637X/713/2/871}, \href
  {https://ui.adsabs.harvard.edu/abs/2010ApJ...713..871W} {713, 871}

\makeatother
\end{thebibliography}

%%%%%%%%%%%%%%%%%%%%%%%%%%%%%%%%%%%%%%%%%%%%%%%%%%

\appendix
\section{VLBA maps and source models}
\label{appendix}

In \autoref{afig:maps}, we show the brightness distributions at 23.8, 5.0, 2.3, 1.8, 1.5 and 1.4~GHz of the quasar 2005$+$403 obtained using BP240A, BG258G, BF025, BG196H and BH222E experiment data. The corresponding map parameters and model fit results are listed in \autoref{atab:maps_info} and \autoref{atab:modelfits}, respectively.

\begin{figure*}
    \centering
    \includegraphics[width=0.45\linewidth]{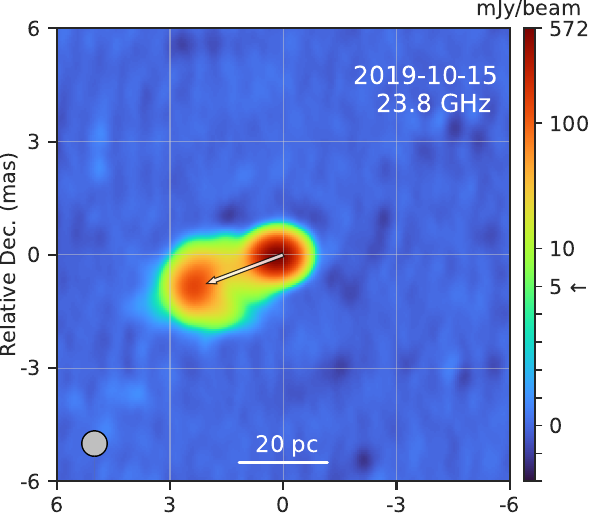}
    \hspace*{0.5cm}
    \includegraphics[width=0.45\linewidth]{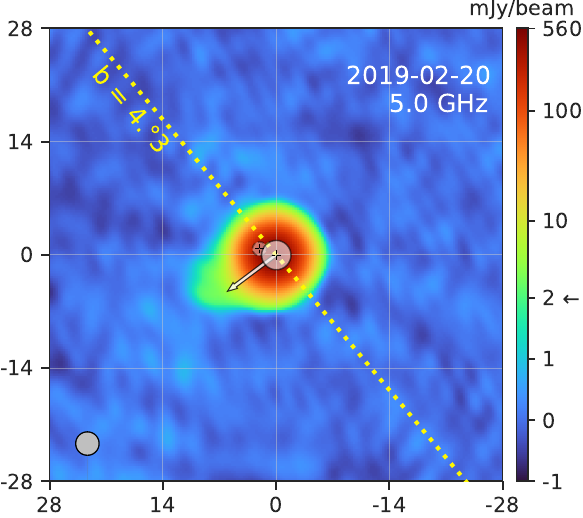}
    \includegraphics[width=0.45\linewidth]{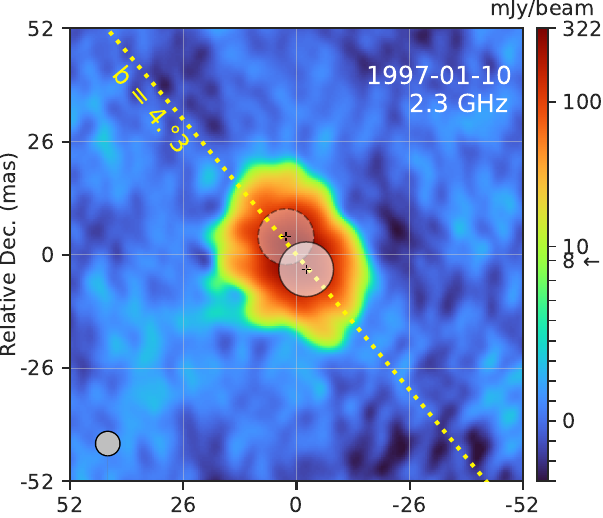}
    \hspace*{0.5cm}
    \includegraphics[width=0.45\linewidth]{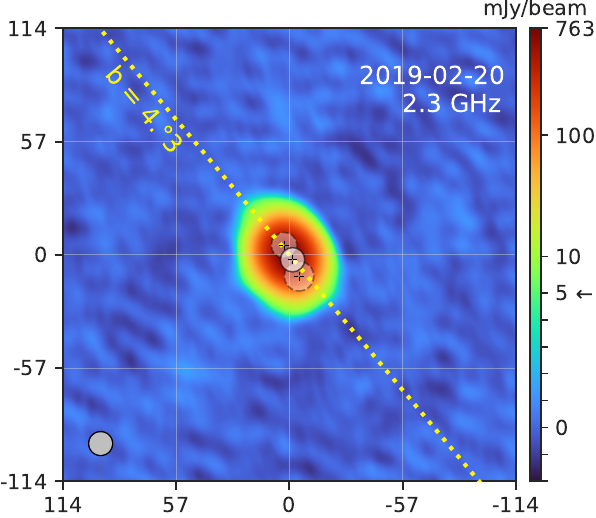}
    \includegraphics[width=0.45\linewidth]{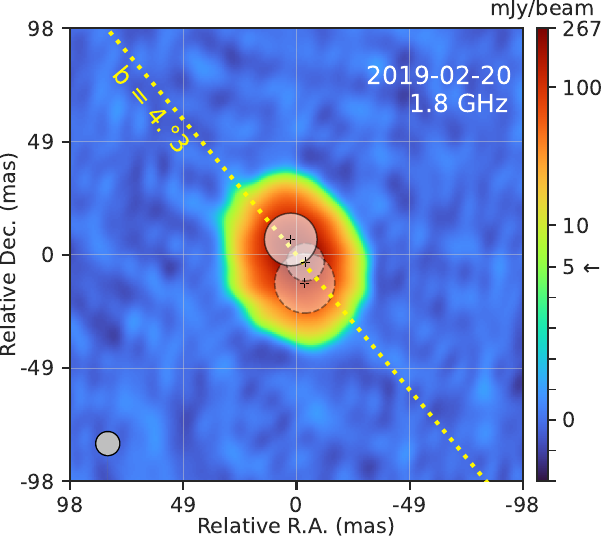}
    \hspace*{0.5cm}
    \includegraphics[width=0.45\linewidth]{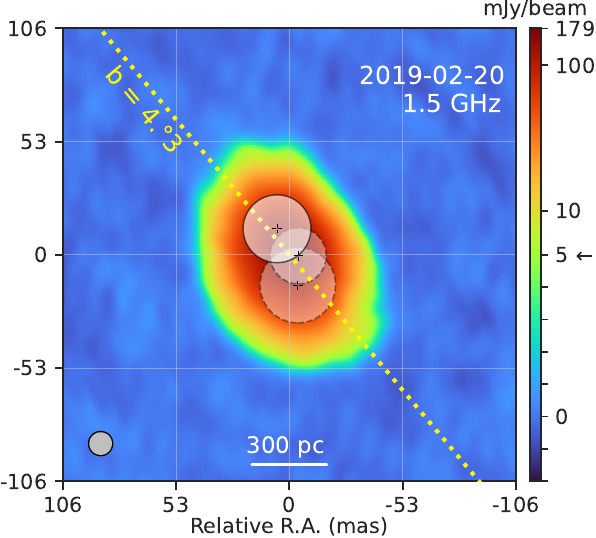}
    \caption{VLBA images of 2005$+$403. Upper left: 23.8~GHz map with the jet at $\mathrm{PA}=110^\circ$. Upper right: 5.0~GHz map with the jet at $\mathrm{PA}=120^\circ$. Middle left and right: 2.3~GHz maps. Bottom left and right: $1.8$ and $1.5$~GHz maps, respectively. The scatter-induced patterns are extended in the direction close to the line of constant Galactic latitude ($b = 4\fdg3$) at $\mathrm{PA}=40\fdg6$ (yellow dotted line). Shaded circles represent the fitted Gaussian components (the brightest feature is shown as solid line circle). To visualize the power of scattering on the observing frequency, the size of each map is matched to the size of the corresponding restoring beam, which FWHM is shown in the lower left corner of each map.}
    \label{afig:maps}
\end{figure*}

\addtocounter{figure}{-1}
\begin{figure*}
    \centering
    \includegraphics[width=0.49\linewidth]{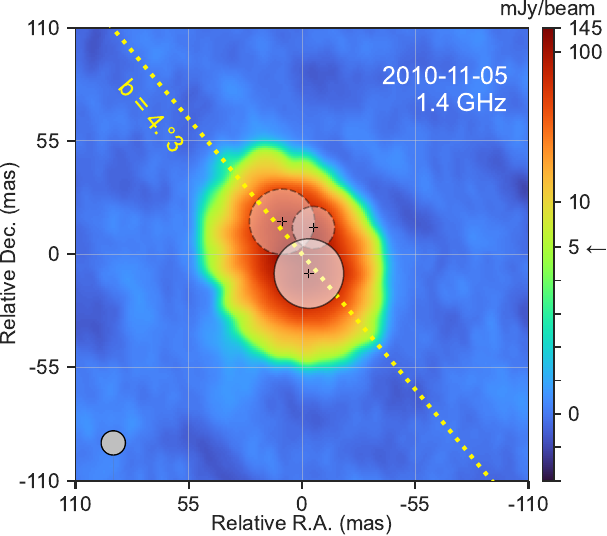}
    %\hspace*{0.5cm}
    \includegraphics[width=0.48\linewidth]{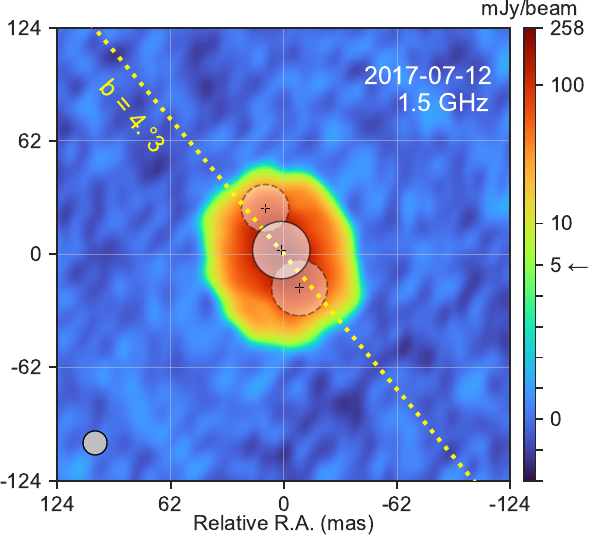}\vspace*{0.3cm}
    
    \includegraphics[width=0.52\linewidth]{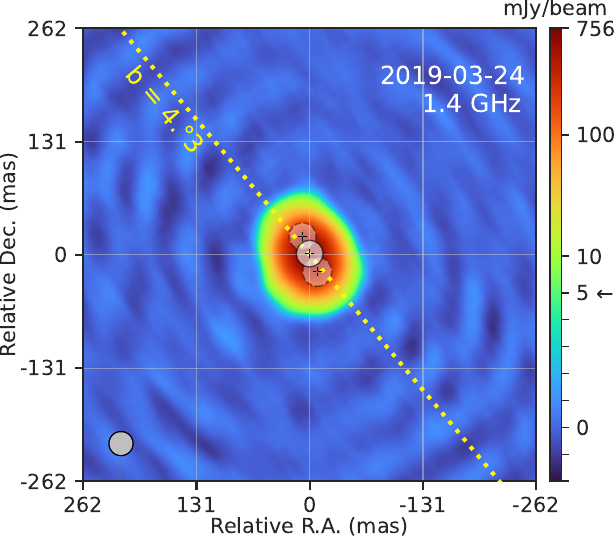}
    \caption{\textit{(continued)}. VLBA images of 2005$+$403 at $1.4$ and $1.5$~GHz. The scatter-induced patterns are extended in the direction close to the line of constant Galactic latitude ($b = 4\fdg3$) at $\mathrm{PA}=40\fdg6$ (yellow dotted line). Shaded circles represent the fitted Gaussian components (the brightest feature is shown as solid line circle). The colorbar scale transits from linear to logarithmic at 5~mJy~beam$^{-1}$. To visualize the power of scattering on the observing frequency, the size of each map is matched to the size of the corresponding restoring beam, which FWHM is shown in the lower left corner of each map.}\vspace*{5cm}
\end{figure*}

%\clearpage
%\newpage
\begin{table*}
	\caption{Map parameters from \autoref{afig:maps}.}
	\centering
	\begin{tabular}{|*{9}{c|}} 
		\hline
		\hline
            Epoch & $\nu$ & $I_\mathrm{peak}$ & $S_\mathrm{tot}$& rms                & Beam  & Pixel size\\
                  & (GHz) & (mJy beam$^{-1}$) & (Jy)            & (mJy beam$^{-1}$)  & (mas) & (mas) \\
            (1)   & (2)   & (3)               & (4)             & (5)                & (6)   & (7)   \\
		\hline
            2019-10-15 & 23.8 & 572.1 & 1.13 & 0.18  & 0.62 &  0.05 \\
            2019-02-20 & 5.0  & 559.7 & 1.59 & 0.21  & 2.70 &  0.3  \\
            1997-01-10 & 2.3  & 322.1 & 2.87 & 1.09  & 5.27 &  0.5  \\    
            2019-02-20 & 2.3  & 763.4 & 2.14 & 0.36  & 11.31&  0.5  \\
            2019-02-20 & 1.8  & 267.3 & 2.05 & 0.30  & 9.77 &  0.8  \\
            2019-02-20 & 1.5  & 178.7 & 2.34 & 0.22  & 10.58&  1.0  \\
            2017-07-12 & 1.5  & 258.2 & 2.28 & 0.45  & 12.32&  1.0  \\
            2010-11-05 & 1.4  & 144.8 & 1.92 & 0.24  & 11.03&  1.0  \\
            2019-03-24 & 1.4  & 756.3 & 2.24 & 0.38  & 26.10&  1.0  \\
		\hline
	\end{tabular}
          \begin{tablenotes}
            \item The columns are as follows: (1) observational epoch; (2) observing central frequency; (3) intensity peak of the map; (4) total flux of the map; (5) the root mean square level of the residuals of the final map; (6) FWHM size of the restoring circular beam; (7) pixel size. 
            \end{tablenotes}
	\label{atab:maps_info}
\end{table*}

\begin{table*}
	\caption{The results of brightness distribution model fitting using circular Gaussian components for the VLBA data used in this work but not listed in \autoref{tab:modelfits}.}
	\centering
	\begin{tabular}{|*{7}{c|}}  
		\hline
		\hline
		Epoch      & $\nu$ & Comp.    & $S$           & $r$            & $\phi$          & $\mathrm{\theta}$\\
                   & (GHz) &          & (Jy)          & (mas)          & ($^\circ$)      & (mas) \\
        (1)        & (2)   & (3)      & (4)           & (5)            & (6)             & (7) \\       
        \hline
        1997-01-10 & 2.3   & core$^*$ & $1.46\pm0.24$ & 0.00           & \ldots          & $12.57\pm2.23$  \\ 
                   &       & sub-c    & $1.41\pm0.24$ & $8.78\pm1.2$   & $31.8\pm7.8$    & $12.84\pm2.41$  \\ 
        \hline
        2010-11-05 & 1.4   & core$^*$ & $1.20\pm0.16$  & 0.00           & \ldots          & $33.76\pm4.86$  \\ 
                   &       & sub-c    & $0.58\pm0.11$ & $28.38\pm3.29$ & $27.4\pm6.6$    & $31.68\pm6.59$  \\ 
                   &       & sub-c    & $0.14\pm0.04$ & $22.44\pm3.20$ & $-5.7\pm8.1$    & $20.58\pm6.39$  \\
        \hline
        2017-07-12 & 1.5   & core     & $1.62\pm0.24$ & 0.00           & \ldots          & $31.49\pm4.91$ \\ 
                   &       & sub-c    & $0.20\pm0.05$ & $24.59\pm3.65$ & $21.0\pm8.4$    & $25.98\pm7.30$ \\ 
                   &       & sub-c    & $0.46\pm0.13$ & $22.92\pm4.78$ & $-154.3\pm11.8$ & $30.58\pm9.57$ \\    
        \hline
        2019-02-20 & 5.0   & core     & $1.53\pm0.08$ & 0.00           & \ldots          & $3.62\pm0.22$ \\ 
                   &       & sub-c    & $0.03\pm0.01$ & $2.21\pm0.38$  & $69.3\pm9.7$    & $1.74\pm0.76$ \\ 
                   &       & jet      & $0.02\pm0.02$ & $5.83\pm1.72$  & $124.9\pm16.4$  & $4.58\pm3.44$ \\ 
        \hline
                   & 2.3   & core     & $1.04\pm0.06$ & 0.00           & \ldots          & $12.23\pm0.98$ \\ 
                   &       & sub-c    & $0.91\pm0.06$ & $8.48\pm0.52$  & $30.2\pm3.5$    & $12.86\pm1.04$ \\ 
                   &       & sub-c    & $0.16\pm0.03$ & $8.93\pm1.54$  & $-158.5\pm9.8$  & $14.75\pm3.08$ \\ 
                   &       & jet      & $0.02\pm0.01$ & $18.35\pm1.74$ & $110.1\pm5.4$   & $6.51\pm3.49$  \\ 
        \hline
                   & 1.8   & core$^*$ & $1.08\pm0.15$ & 0.00           & \ldots          & $22.80\pm3.34$ \\ 
                   &       & sub-c    & $0.54\pm0.11$ & $19.79\pm2.76$ & $-162.2\pm7.9$  & $25.95\pm5.52$ \\ 
                   &       & sub-c    & $0.43\pm0.07$ & $11.67\pm1.47$ & $-147.1\pm7.2$  & $16.30\pm2.94$ \\ 
        \hline
                   & 1.5   & core$^*$ & $0.97\pm0.17$ & 0.00           & \ldots          & $31.75\pm5.94$ \\ 
                   &       & sub-c    & $0.88\pm0.17$ & $28.13\pm3.65$ & $-159.7\pm7.4$  & $35.28\pm7.29$ \\ 
                   &       & sub-c    & $0.50\pm0.10$ & $16.14\pm2.77$ & $-141.9\pm9.7$  & $26.23\pm5.54$ \\      
        \hline

        2019-03-24 & 1.4   & core     & $1.41\pm0.11$ & 0.00           & \ldots          & $30.88\pm3.06$ \\ 
                   &       & sub-c    & $0.47\pm0.07$ & $22.71\pm2.97$ & $-157.5\pm7.4$  & $33.48\pm5.94$ \\ 
                   &       & sub-c    & $0.36\pm0.05$ & $21.64\pm2.71$ & $22.3\pm7.1$   & $31.36\pm5.43$ \\
                   
        \hline
        2019-10-15 & 23.8  & core     & $0.34\pm0.03$ & 0.00           & \ldots          & $0.17\pm0.05$  \\ 
                   &       & jet      & $0.33\pm0.03$ & $0.28\pm0.03$  & $95.0\pm6.4$    & $0.20\pm0.06$  \\ 
                   &       & jet      & $0.22\pm0.07$ & $1.96\pm0.22$  & $108.2\pm6.3$   & $1.34\pm0.43$  \\ 
                   &       & jet      & $0.16\pm0.02$ & $2.48\pm0.05$  & $109.2\pm1.1$   & $0.40\pm0.09$  \\ 
                   &       & jet      & $0.07\pm0.02$ & $0.55\pm0.07$  & $102.2\pm7.3$   & $0.18\pm0.14$  \\  
        \hline
	\end{tabular}
         \begin{tablenotes}
            \item The columns are as follows: (1) observing epoch; (2) central observing frequency; (3) type of a component, where 'sub-c' means 'sub-component' of the core; (4) measured flux density of a component; (5) radial distance of a component relative to the core; (6) position angle of a component relative to the core; (7) FWHM of the measured size of a component.
            \item $^*$ VLBI core position is uncertain. 

        \end{tablenotes}
	\label{atab:modelfits}
\end{table*}

% Don't change these lines
\bsp % typesetting comment
\label{lastpage}
\end{document}